\documentclass[a4paper,fleqn,usenatbib]{mnras}

\usepackage[T1]{fontenc}
\usepackage{ae,aecompl}
\usepackage{graphicx}
\usepackage{amsmath}
\usepackage{amssymb}
\usepackage{xcolor}
\usepackage{xspace}
\usepackage{txfonts}
\usepackage{ulem}

\newcommand{\Te}{{T_{\rm e}}}
\newcommand{\ye}{{y_{\rm e}}}
\newcommand{\kB}{{k_{\rm B}}}
\newcommand{\me}{{m_{\rm e}}}
\newcommand{\sigT}{{\sigma_{\rm T}}}

\newcommand{\expf}[1]{{\rm e}^{#1}}

\newcommand{\Rtwohc}{R_{\rm 200c}}
\newcommand{\Mtwohc}{M_{\rm 200c}}
\newcommand{\Rtwohm}{R_{\rm 200m}}
\newcommand{\Mtwohm}{M_{\rm 200m}}
\newcommand{\Rfivehc}{R_{\rm 500c}}
\newcommand{\Mfivehc}{M_{\rm 500c}}
\newcommand{\Rfivehm}{R_{\rm 500m}}

\newcommand{\Rvir}{R_{\rm vir}}
\newcommand{\Mvir}{M_{\rm vir}}
\newcommand{\fgas}{f_{\rm gas}}
\newcommand{\Mgas}{M_{\rm gas}}

\newcommand{\Bah}{\textsc{Bahamas}}
\newcommand{\Mac}{\textsc{Macsis}}
\newcommand{\Mag}{\textsc{Magneticum}}
\newcommand{\Tng}{\textsc{TNG}}
\newcommand{\The}{\textsc{The300}}
\newcommand{\BaM}{\textsc{Bahamas+Macsis}}

\newcommand{\msol}{\ensuremath{\, {\rm M}_\odot}}

\newcommand{\Tm}{T_{\rm m}}
\newcommand{\Ty}{T_{\rm y}}
\newcommand{\Tsl}{T_{\rm sl}}
\newcommand{\slog}{\sigma_{\log_{10}T}}

\newcommand{\topic}[1]{\vspace{0.75mm}{\noindent{\it #1}}}

\title[SZ temperature scalings]{A multi-simulation study of relativistic SZ temperature scalings in galaxy clusters and groups}

\author[E. Lee et al.]{
Elizabeth Lee$^{1}$\thanks{E-mail: elizabeth.lee-2@postgrad.manchester.ac.uk},
Dhayaa Anbajagane$^{2,3}$,
Priyanka Singh$^{4,5}$,
Jens Chluba$^{1}$,
Daisuke Nagai$^{4,5}$,
Scott T. Kay$^{1}$,
\newauthor
Weiguang Cui$^{6, 7}$,
Klaus Dolag$^{8, 9}$
and Gustavo Yepes$^{6}$
\\
$^{1}$Jodrell Bank Centre for Astrophysics, Department of Physics and Astronomy, The University of Manchester, Manchester M13 9PL, UK \\
$^{2}$Department of Astronomy and Astrophysics, University of Chicago, Chicago, IL 60637, USA \\
$^{3}$Kavli Institute for Cosmological Physics, University of Chicago, Chicago, IL 60637, USA \\
$^{4}$Department of Physics, Yale University, New Haven, CT 06520, USA \\
$^{5}$Yale Center for Astronomy \& Astrophysics, 52 Hillhouse Avenue, New Haven, CT 06511, USA \\
$^{6}$Departamento de F\'isica Te\'{o}rica, M\'{o}dulo 15, Facultad de Ciencias, Universidad Aut\'{o}noma de Madrid, 28049 Madrid, Spain\\
$^{7}$Institute for Astronomy, University of Edinburgh, Royal Observatory, Edinburgh EH9 3HJ, United Kingdom\\
$^{8}$Universit\"ats-Sternwarte, Fakult\"at für Physik, Ludwig-Maximilians-Universit\"at M\"unchen, Scheinerstr.1, 81679 M\"unchen, Germany \\
$^{9}$Max-Planck-Institut f\"ur Astrophysik, Karl-Schwarzschild-Straße 1, 85741 Garching, Germany\\
}

\date{\vspace{-3mm}
Accepted XXX. Received YYY; in original form ZZZ}

\pubyear{2022}

\begin{document}
\label{firstpage}
\pagerange{\pageref{firstpage}--\pageref{lastpage}}
\maketitle

\begin{abstract}
The Sunyaev-Zeldovich (SZ) effect is a powerful tool in modern cosmology. With future observations promising ever improving SZ measurements, the relativistic corrections to the SZ signals from galaxy groups and clusters are increasingly relevant. 
As such, it is important to understand the differences between three temperature measures: (a) the average relativistic SZ (rSZ) temperature, (b) the mass-weighted temperature relevant for the thermal SZ (tSZ) effect, and (c) the X-ray spectroscopic temperature. In this work, we compare these cluster temperatures, as predicted by the {\sc Bahamas} \& {\sc Macsis}, {\sc IllustrisTNG}, {\sc Magneticum}, and {\sc The Three Hundred Project} simulations.
Despite the wide range of simulation parameters, we find the SZ temperatures are consistent across the simulations. We estimate a $\simeq 10\%$  level correction from rSZ to clusters with $Y\simeq10^{-4}$~Mpc$^{-2}$. Our analysis confirms a systematic offset between the three temperature measures; with the rSZ temperature $\simeq 20\%$ larger than the other measures, and diverging further at higher redshifts. We demonstrate that these measures depart from simple self-similar evolution and explore how they vary with the defined radius of haloes.
We investigate how different feedback prescriptions and resolutions affect the observed temperatures, and discover the SZ temperatures are rather insensitive to these details.
The agreement between simulations indicates an exciting avenue for observational and theoretical exploration, determining the extent of relativistic SZ corrections. We provide multiple simulation-based fits to the scaling relations for use in future SZ modelling.
\end{abstract}


\section{Introduction}
In recent years, the Sunyaev-Zeldovich (SZ) effect \citep{Zeldovich1969,Sunyaev1970} has become a powerful tool for studying the warm-hot gas in the universe \citep[][for a recent review]{Planck2014XX, Pratt2019}. Advances in high-resolution SZ observations promise to improve our understanding of thermodynamic and kinetic properties of the warm hot universe \citep[see][for a recent review]{Mroczkowski2019}. The thermal and kinematic SZ effects have been widely used to measure thermal pressure \citep[e.g.,][]{Mroczkowski2009,Planck2013V} and the peculiar velocity of clusters \citep[e.g.,][]{Sayers2019}, with applications for both astrophysical and cosmological studies.

A promising next step lies in the study of the relativistic SZ (rSZ) corrections \citep{Sazonov1998, Nozawa2006, Chluba2012} which depend on the cluster temperature. Despite many attempts in single cluster observations \citep[e.g.,][]{Hansen2002, Hansen2004, Zemcov2012, Prokhorov2012, Chluba2013, Butler2021} and stacking analyses \citep[e.g.,][]{Hurier2016,Erler2018}, the detection significance of the rSZ effect remains low, and to date direct constraints on the temperatures relevant for SZ measurements have been limited.
However, the continued advances in both angular resolution and sensitivities, as well as upcoming high-resolution, multi-frequency SZ observations, such as CCAT-Prime \citep{Stacey2018}, NIKA2 \citep{NIKA2updated}, TolTEC \citep{TolTEC}, and The Simons Observatory \citep{TheSimonsObservatoryCollaboration2018}, promise to measure ICM temperature using the rSZ effect for individual galaxy clusters and stacked samples of groups, especially in high-redshift ($z\gtrsim 1$) objects \citep{Morandi2013} without any aid of X-ray observations.
Moreover, even when the cluster temperatures cannot be measured with sufficient statistical certainty, the omission of rSZ effects leads to an underestimation of the Compton-$y$ parameter, especially for the most massive clusters. This can impact ongoing and future cosmological analyses, showing that it is imperative to identify ways of including rSZ in the signal modeling \citep{Remazeilles2018, Remazeilles2019}.

Modern hydrodynamical cosmological simulations have become an indispensable tool for understanding the ICM structures and evolution and their impact on SZ observables \citep[e.g.,][]{Nagai2006,Nagai2007b,Battaglia2012,Kay2012,Pike2014,Yu2015,Planelles2017,LeBrun2017,Henden2018,Henden2019,Pop2022a,Pop2022b}. However, these simulations are also known to exhibit significant variations among different hydrodynamic codes \citep{Rasia2014, Sembolini2016}, which gives rise to differences in a variety of cluster observables, such as the hydrostatic mass derived from X-ray mocks \citep{Rasia2006,Nagai2007a}.  
It is hence important to develop complementary approaches for measuring the ICM temperature based on SZ effect observations. The rSZ effect with upcoming SZ observations promises to provide insights into the thermodynamic structure and evolution \citep{Lee2020}.

Due to the inherent variation of temperature within clusters and groups, both being neither isothermal nor spherically distributed \citep[e.g.,][]{Nagai2007b, Vikhlinin2009cosmoparams}, the temperature used in cluster analysis depends on the context in which they are applied. This has been discussed in detail for SZ measurements alongside X-ray observations \citep[e.g.,][]{Pointecouteau1998,Hansen2004,Kay2008,Lee2020}.
While simulations encompass many variations, it is exciting to explore what can be learnt from common behaviour between simulations. In particular, we find the SZ temperatures demonstrate consistency between simulations, allowing for predictions of the rSZ corrections that can be detected observationally, in addition to determining what observations may tell us about the underlying physics.

In this work, we study a sample of clusters and groups extracted from 4 different hydrodynamical simulations in a follow-up to earlier work aiming at establishing detailed rSZ temperature-mass relations \citep{Lee2020}. In particular, we use the hydrodynamical simulations $\Bah$ \citep{McCarthy2017, McCarthy2018};  $\Mac$ \citep{Barnes2017}; {\sc The Three Hundred Project} \citep{Cui2018}; {\sc Magneticum Pathfinder} \citep{Hirschmann2014, Bocquet2016}; and {\sc IllustrisTNG} \citep{Springel2018,Pillepich2018FirstResults,Nelson2018,Marinacci2018, Naiman2018}. These simulations provide large samples of groups and clusters 
over 5 redshifts between $z=0$ to $z=1.5$. In each simulation, we study three different temperature measures: the spectroscopic-like temperature \citep{Mazzotta2004}; the mass-weighted temperature, which is closely related to the Compton-$y$ parameter; and finally the y-weighted temperature, which is a close approximation for the averaged rSZ temperature, needed to account for the rSZ corrections \citep{Hansen2004, Remazeilles2019}.

Our work is an extension to a growing body of literature that uses ensembles of the latest hydrodynamical simulations to estimate theoretical uncertainties in cluster scaling relations. These uncertainties arise from our lack of knowledge of the true astrophysical mechanisms at play in our Universe. Either some, or all of the simulations we use in this work have been used previously to marginalize over astrophysics models and estimate scaling relation uncertainties for different cluster properties, such as the thermal gas pressure \citep{Lim2021}, the central and satellite galaxy stellar properties \citep{Anbajagane2020}, and the dark matter and satellite galaxy velocity dispersions \citep{Anbajagane2022}. Here we extend the discussion to the modeling of the rSZ effect.

This paper is structured as follows. In Section~\ref{sec:Temperature_Weightings}, we discuss the mathematical background behind each temperature measure, alongside some of the formalism we use. We discuss the simulation and our halo samples in Section~\ref{sec:Simulations}. In Section~\ref{sec:TS}, we examine the behaviour of the cluster-averaged temperatures in each simulation and how they vary with mass, redshift, radius, Compton-$y$ parameter, and other temperature measures. Section~\ref{sec:cross_sim_fits} contains an exploration of the cross-simulation averaged properties and fits. We discuss applications of our results in Section~\ref{sec:applications}. We summarize our findings in Section~\ref{sec:Conclusion}.

\vspace{-4mm}
\section{Theory}
\label{sec:Temperature_Weightings}
In this paper, we follow much of the same formalism introduced in \cite{Lee2020}; however, for completeness, we briefly review it here. We start by examining each temperature measure. We then briefly discuss observed temperature scaling relations, and finally mention a number of specifics relevant to calculating temperatures consistently within simulations.

In general, when we consider groups and clusters within simulations, we are in fact calculating the signal from a sphere co-located with the dark matter halo, of a certain radius, $R_\Delta$. These radii are defined as the radii containing a certain averaged density, and thus mass $M_\Delta$, as will be explored further in Sect.~\ref{sec:MvR_theory}. This is a reflection of current SZ observations which often only have the angular resolution or sensitivities to measure spherically averaged temperature structures. As such most observationally relevant quantities are integrated over a sphere (or a projected circle on the sky).

These haloes evidently do not have constant densities nor temperatures throughout the observed sphere, so any observed temperature will instead be an averaged quantity. These averages are, however, always defined by the weighting procedure, which depends on the observable at hand. We can therefore define
\begin{equation}
\langle T\rangle \equiv \frac{\int wT\,{\rm d}V}{\int w\,{\rm d}V},
\end{equation}
where $w$ represents the weighting. As discussed later in this section, for the mass-weighted, $y$-weighted and spectroscopic-like temperatures one respectively has $w=n$, $w=nT$ and $w=n^2T^{-\alpha}$ with $\alpha\simeq0.75$ and $n$ and $T$ are the electron density and temperature, respectively.

\vspace{-4mm}
\subsection{SZ Temperatures}
\topic{Compton-$y$ parameter}: The classical thermal SZ (tSZ) signal has an amplitude proportional to the Compton-$y$ parameter. This Compton-$y$ parameter is proportional to the integrated electron pressure, $P_{\rm e}$:
\begin{equation}
\label{eq:yparameter}
    y \equiv \int \frac{\kB \Te}{\me c^2}\,{\rm d}\tau = \frac{\sigT}{\me c^2}\int P_{\rm e}\,{\rm d}l = 
    \frac{\sigT \kB}{\me c^2}\int n_{\rm e}\Te\, {\rm d}l.
\end{equation}
Here, $\tau$ is the Thomson scattering optical depth; ${\rm d}l$ parametrizes the integral along the line of sight. All the other constants have their usual meaning.

The non-relativistic tSZ effect is then written as \citep{Zeldovich1969}
\begin{equation}
\label{eq:non-rel-tSZ}
\Delta I_\nu \equiv I_0\, y\, g(x) = I_0\, y \frac{x^4 \expf{x}}{(\expf{x}-1)^2}\left(x\frac{\expf{x}+1}{\expf{x}-1}-4\right),
\end{equation}
where $g(x)$, the tSZ spectral function, is defined implicitly. We also use the dimensionless frequency $x=h\nu/\kB T_{\rm CMB}$, with the present-day CMB temperature $T_{\rm CMB}=2.7255\,{\rm K}$. The intensity normalisation constant is $I_0 = 270.33 (T_{\rm CMB}/2.7255{\rm K})^3$~MJy/sr.

\vspace{2mm}
\topic{Mass-weighted temperature}:
Eq.~\eqref{eq:yparameter} and \eqref{eq:non-rel-tSZ} directly motivate the use of a mass-weighted or $\tau$-weighted temperature:
\begin{equation}
    \Tm \equiv \frac{\int \Te\,{\rm d}m}{\int {\rm d}m} = \frac{\int n_{\rm e}\Te\,{\rm d}V}{\int n\,{\rm d}V},
\end{equation}
with $m$ the mass of the electron gas. Hence the volume integrated Compton-$y$ parameter, $Y$, is
\begin{equation}
    Y = \frac{\sigT \kB}{\me c^2}\int n_{\rm e}\Te\, {\rm d}V \propto \Mgas\Tm
\end{equation}
where $\Mgas$ is the total gas mass within the halo. From observations, this temperature measurement can be estimated by combing tSZ and X-ray measurements, where the latter is used to obtain a mass/$\tau$ estimate. However, since the X-ray temperature does not have the same weighting (see below) and because non-thermal pressure contributions can affect the inference, this leads to a mass bias \citep[e.g.,][]{Arnaud2005, Nagai2007a, Battaglia2012SZphysics, Nelson2012, Nelson2014b, Shi2016}. It is also worth noting that a similar method can be used by combining measurements from the kinematic SZ and tSZ effect which would minimise this bias \citep[e.g.,][]{Lim2020}.

\vspace{2mm}
\topic{$y$-weighted temperature}: The relativistic corrections to the SZ effect lead to a temperature-dependent modification to the spectral function which can be accurately modelled using {\tt SZpack} \citep{Chluba2012,Chluba2013}. This implies that we should write for the relativistically-corrected SZ signal, $\Delta I_\nu = I_0 y f(x,\Te)$, in an isothermal cluster. However, since the temperature varies within each cluster, we can write an expansion about an arbitrary temperature pivot $\bar{T}$ \citep{Chluba2013,Remazeilles2018}. To second order in $\Delta T=\Te-\bar{T}$, we have
\begin{equation}
    \frac{\Delta I}{I_0}\simeq yf(x,\bar{T})+y^{(1)}(\bar{T})f^{(1)}(x,\bar{T})+\frac{1}{2}y^{(2)}(\bar{T})f^{(2)}(x,\bar{T}),
\end{equation}
where $f^{(k)}=\partial_T^{k}f(x,\Te)$ and $y^{(k)}=\langle(T-\bar{T})^k \ye\rangle$. Here, $\langle X\rangle$ is indicating the cluster averaged value, and we are now using a `local' $y$, $\ye\propto n_{\rm e}\Te$. This motivates a relativistic temperature, which removes the first-order correction (i.e., $y^{(1)}(\bar{T})=0$) and we call the $y$-weighted temperature
\begin{equation}
    \Ty \equiv \frac{\int \ye \Te\,{\rm d}V}{\int \ye\,{\rm d}V} =\frac{\int n_{\rm e} \Te^2\,{\rm d}V}{\int n_{\rm e}\Te\,{\rm d}V}. 
\end{equation}
We note that there will still be contributions from higher order temperature terms, however in \citet{Lee2020}, these are determined to lead to $\simeq0.1\%$ corrections to the SZ signal peak, and as such we neglect the further study of them here. We find that this $y$-weighted temperature is systematically higher than the mass-weighted and X-ray temperatures, indicating that, especially for the largest clusters in the Universe, the relativistic corrections will be relevant and may bias cosmological inferences if not modelled.

\vspace{-4mm}
\subsection{X-ray Temperatures}
X-ray emission, from hot clusters ($\kB \Te\gtrsim 3$~keV), is dominated by bremstrahllung radiation within the ICM, while cooler clusters have increasing contributions from emission lines.
However, in this hot regime, temperature weightings have been determined by fitting the X-ray spectra with a thermal emission model \citep{Mazzotta2004, Vikhlinin2006}. This has led to the \topic{spectroscopic-like temperature},
\begin{equation}
    \Tsl \equiv \frac{\int n_{\rm e}^2\Te^{1-\alpha}\,{\rm d}V}{\int n_{\rm e}^2\Te^{-\alpha}\,{\rm d}V}
\end{equation}
with $\alpha=0.75$ for high temperatures ($\kB \Te>3.5$~keV), which matches well with observations from both {\it Chandra} and {\it XMM-Newton}. X-ray temperatures have also been calibrated within simulations, to determine the differences between different X-ray temperatures and to confirm the weighting for $\Tsl$ \citep[e.g.,][]{Rasia2014}.

\vspace{-4mm}
\subsection{Halo definition and redshift dependence}
\label{sec:MvR_theory}
\topic{Radius definitions}: As previously mentioned, to define the extent of groups and clusters within our simulations, we consider spheres of a given radius, co-located with the halo, i.e., centred on the minimum of potential. Then a halo of radius $R_\Delta$, is defined so that it contains the mass $M_\Delta$.

In this work, we use five different radii, in particular, $\Rfivehc$, $\Rtwohc$, $\Rfivehm$, $\Rtwohm$, and $\Rvir$. For $R_{\rm \Delta c}$, these are defined in terms of an overdensity of $\Delta$ times the critical density of the universe, $\rho_{\rm crit}$. i.e.,
\begin{equation}
    M_{\rm \Delta c} = \frac{4\pi}{3}R_{\rm \Delta c}^3 \rho_{\rm crit}(z) \Delta,
\end{equation}
while the $R_{\rm \Delta m}$ are defined in terms of overdensities of $ \rho_{\rm mean}\Delta$, the mean density of the universe. The critical density and mean density are defined as,
\begin{equation} \begin{split}
    \rho_{\rm crit} &\equiv \frac{3H_0^2}{8\pi G}\left[\Omega_m(1+z)^3+\Omega_\Lambda\right] = \frac{3H_0^2}{8\pi G}E^2(z)\\
    \rho_{\rm mean} &\equiv \frac{3H_0^2}{8\pi G}\left[\Omega_m(1+z)^3\right]
\end{split} \end{equation}
for a flat $\Lambda$CDM universe. We have also here implicitly defined $E(z)$, which describes the redshift evolution of the Hubble parameter.
The virial radius $\Rvir$ is defined by $\rho_{\rm crit}$ and $\Delta_{\rm vir}$ calculated using the approximation in \cite{Bryan1998}, that is, $\Delta_{\rm vir}\approx102$ at $z=0$.

In this work, we predominantly define our clusters as being within the radius $\Rtwohc$, a common proxy for the observational radii used in many SZ calculations.
We note that here we are always using the {\it true} (simulation) mass, rather than any proxy for the observed mass (i.e., the hydrostatic mass), which could introduce observational biases \citep[e.g.,][for a recent review]{Nagai2007a,Lau2009,Lau2013,Nelson2012,Nelson2014a,Shi2016,Biffi2016,Barnes2017, Pratt2019}. As such for direct comparison with any observational quantities, the associated mass bias must be considered, however a detailed mass calibration based on mocks for specific observations is beyond the scope of this work.

\topic{Redshift dependence}: As the critical density has a redshift dependence, we can accordingly expect an evolution of the cluster properties. Simple geometric consideration and assumption of isothermality within the virialized sphere would lead to a temperature dependence akin to
\begin{equation}
    T_\Delta \propto M_\Delta/R_\Delta.
\end{equation}
Then, by using the $\rho_{\rm crit}$ redshift dependence, we would expect a self-similar evolution of
\begin{equation} \begin{split}
    M_\Delta &\propto E^2(z) R_\Delta^3,\\
    T_\Delta &\propto E^{2/3}(z) M_\Delta^{2/3}.
\end{split} \end{equation}
As such we can consider any deviation from this evolution as being inherent evolution of the temperatures, due to non-gravitational processes in the clusters themselves. In this work, we use $E(z)$ according to the cosmology for each simulation independently as there are substantial variations, as discussed in the next section.

\vspace{-4mm}
\section{Simulations}
\label{sec:Simulations}
\begin{table*}
	\centering
	\begin{tabular}{l|c|c|c|c|c|l}
		Simulation & $L$ [Mpc] & $N$ & $\epsilon_{\rm \,DM}^{z=0} \,\rm [kpc]$ & $m_{\rm DM} \, [\msol]$ & $m_{\rm gas} \, [\msol]$ & Calibration \\
        \hline
            $\Bah$ & $596$ & $2 \times 1024^3$ & $5.96$ & $6.7 \times 10^9$ & $1.2 \times 10^9$ & GSMF, CL $\fgas$ \\
            $\Mac$ & ---   & ---    & $5.90$ & $6.5 \times 10^9$ & $1.2 \times 10^9$ & GSMF, CL $\fgas$ \\
            $\The$ & ---   & ---    & $9.6$  & $1.9 \times 10^9$ & $3.5 \times 10^8$ & See \citet{Cui2018} \\
            $\Mag$ & $500$ & $2 \times 1584^3$& $5.33$ & $9.8 \times 10^8$ & $2.0 \times 10^8$ & SMBH, Metals, CL $\fgas$ \\
            $\Tng$ & $303$ & $2 \times 2500^3$ & $1.48$ & $5.9 \times 10^7$ & $1.1 \times 10^7$ & See \citet{Pillepich2018SimulatingGalaxyFormation} \\
            \hline
    \end{tabular}
    \caption{Simulation characteristics for each sample. Table adapted from \citet[Table 1]{Anbajagane2022}. From left to right, we show: (i) simulation, (ii) comoving box size, (iii) the total particle count, (iv) force softening scale, (v) mass of DM particles, (vi) mass of gas particles and (vii) empirical sources used for tuning sub-grid parameters of each simulation, which consist of the Galaxy Stellar Mass Function (GSMF), Supermassive Black Hole scaling (SMBH), Metallicity scaling (Metals), and cluster hot gas mass fraction $<R_{500c}$ (CL $\fgas$). We do not quote the box sizes or particle numbers for $\Mac$ and $\The$ as they are zoom-in simulations.}
	\label{tab:SimulationParams}
\end{table*}
\begin{table}
\caption{Cosmological parameters used in each of the simulations.}
\begin{tabular}{l |@{}c |@{}c |@{}c |@{}c |@{}c |@{}c |@{}c }
    \hline
    Simulation & $\Omega_\Lambda$ & $\Omega_\mathrm{m}$ & $\Omega_\mathrm{b}$ & $\sigma_8$ & $n_\mathrm{s}$ &$h^\dagger$\\
    \hline
    $\Bah$ & 0.6825 & 0.3175 & 0.0490 & 0.8340 & 0.9624 & 0.6711 \\
    $\Mac$ & 0.6930 & 0.3070 & 0.0482 & 0.8288 & 0.9611 & 0.6777 \\
    $\The$ & 0.6930 & 0.3070 & 0.0480 & 0.8230 & 0.9600 & 0.6780 \\
    $\Mag$ & 0.7280 & 0.2720 & 0.0457 & 0.8090 & 0.9630 & 0.7040 \\
    $\Tng$ & 0.6911 & 0.3089 & 0.0486 & 0.8159 & 0.9667 & 0.6774 \\
    \hline \noalign{\smallskip}
    \multicolumn{7}{l}{$^\dagger$ where $h \equiv H_0/(100 \,\mathrm{km\,s}^{-1}\,\mathrm{Mpc}^{-1})$ }
\end{tabular}
\label{tab:SimulationCosmology}
\end{table}

We use four samples of haloes from the following simulations: (i) a superset of $\Bah$ and $\Mac$, (ii) {\sc The Three Hundred Project}, (iii) {\sc Magneticum Pathfinder}, and (iv) {\sc IllustrisTNG}. A summary of the simulation parameters in Table~\ref{tab:SimulationParams} and the cosmological properties of each simulation can be found in Table~\ref{tab:SimulationCosmology}.

We take haloes within each simulation with $\Mfivehc >10^{13}$~$\msol$ and consider the redshifts $z \simeq \{0.0, 0.25, 0.5, 1.0, 1.5\}$. The population counts at each redshift can be found in Table~\ref{tab:SimulationPopulations}.
In each simulation, we calculate our properties using a temperature cut -- that is, only considering particles with temperatures, $\Te \geq 10^{5.2}$~K. Moreover, when calculating the {\it spectroscopic-like} temperature, we use a core excision procedure as has been shown to make X-rays a better proxy for mass \citep[e.g.,][]{Pratt2009}, and exclude all particles with $r<0.15\,\Rfivehc$. We also note in Appendix~\ref{app:relaxation}, that using a relaxed subsample does not seem to meaningfully change our results at $z=0$, so this is not further examined here.

We will now briefly discuss the specifics of each simulation. However, a detailed discussion of the differences in the feedback models and physics used in each of these simulations is beyond the scope of this paper. The wide breadth of differences between these simulations makes it very difficult to infer which factor of each simulation might lead to any observed similarities or differences. Where it is possible -- e.g., for feedback models (Appendix~\ref{sec:Res_Feedback}) and resolution (Appendix~\ref{sec:Res_Resolution}) -- we examine the differences within a suite of simulations where only these factors were affected.

\begin{table*}
    \begin{tabular}{l|cc|cc|cc|cc|cc}
        Nominal redshift & \multicolumn{2}{c|}{$z=0.00$} & \multicolumn{2}{c|}{$z=0.25$} & \multicolumn{2}{c|}{$z=0.50$} & \multicolumn{2}{c|}{$z=1.00$} & \multicolumn{2}{c}{$z=1.50$} \\
         & $z_{\rm sim}$ & $N_{\rm haloes}$ & $z_{\rm sim}$ & $N_{\rm haloes}$ & $z_{\rm sim}$ & $N_{\rm haloes}$ & $z_{\rm sim}$ & $N_{\rm haloes}$ & $z_{\rm sim}$ & $N_{\rm haloes}$ \\ \hline
        $\BaM$ & 0.00 & 21361 & 0.25/0.24 & 19610 & 0.50/0.46 & 17016 & 1.00 & 10928 & 1.50/1.56 & 5750 \\
        $\The$ & 0.00 & 8465 & 0.25 & 8433 & 0.49 & 8300 & 0.99 & 7349 & 1.48 & 5627 \\
        $\Mag$ & 0.066 & 10429 & 0.25 & 9812 & 0.52 & 8265 & 0.96 & 5659 & 1.47 & 3080 \\
        $\Tng$ & 0.00 & 2548 & 0.24 & 2333 & 0.50 & 2015 & 1.00 & 1290 & 1.50 & 700
    \end{tabular}
    \caption{The number of haloes with $\Mfivehc>10^{13}$~$\msol$ within each simulation, at each considered redshift. The nominal redshift is the how we will refer to each sample, while $z_{\rm sim}$ indicates the `true' simulation redshift used. For the $\BaM$ sample, where the redshifts are recorded as `x/y', the first number refers to $z_{\rm sim}$ in the $\Bah$ simulation, while the second is that in the $\Mac$ simulation.}
    \label{tab:SimulationPopulations}
\end{table*}

\vspace{-4mm}
\subsection{Bahamas and Macsis}
We use a supersample of $\Bah$ \citep{McCarthy2017, McCarthy2018} and its zoom-in counterpart $\Mac$ \citep{Barnes2017}, which we will here refer to as $\BaM$, or in figures simply $\Bah$+.
$\Bah$ is a calibrated version of the model used in the cosmo-OWLS simulations \citep{LeBrun2014}. Both $\Bah$ and $\Mac$ were run using a version of the smoothed particle hydrodynamics (SPH) code {\tt GADGET-3} last publicly discussed in \cite{Springel2005}, but modified as part of the OWLS Project \citep{Schaye2010}. The $\Mac$ simulation was developed to extend the $\Bah$ sample to higher mass haloes. It is a sample of 390 clusters, generated through a zoomed simulation from a large Dark Matter only (DMO) simulation -- a periodic cube with a side length of 3.2~Gpc. Individual separate volumes including high Friends-of-Friends (FoF) mass (clusters with $M_{\rm FoF}>10^{15}$~$\msol$) were then re-simulated with a full hydrodynamical prescription aligned with the $\Bah$ simulation. Haloes within the two simulations are then identified by a FoF algorithm, and subhaloes with the {\sc Subfind} algorithm \citep{Springel2001, Dolag2009}. 

To form this supersample, all the haloes with the relevant masses from both simulations are used. The differences between cosmologies are deemed to have minimal effects in general, however, when considering the redshift variation, $E(z)$ is calculated with the relevant cosmology for $\Mac$ and $\Bah$ separately. It is also worth noting, that unlike {\sc The Three Hundred Project}, $\Mac$ only provides 390 massive clusters, and no additional lower mass clusters.

\vspace{-4mm}
\subsection{The Three Hundred Project}
{\sc The Three Hundred Project}, here $\The$, \citep{Cui2018} comprises massive haloes formed within 324 spherical regions each of 22 Mpc radius and each centred on the most massive clusters at redshift zero as identified in the {\sc MultiDark Planck 2} N-body, DMO simulation \citep{Klypin2016}, which has a cube of side length $1.476$~Gpc with $3840^3$ dark matter particles and used the {\tt L-GADGET-2} solver. 
The haloes in {\sc MultiDark Planck 2} were identified using the {\sc Rockstar} halo finder \citep{Behroozi2013}.
These 324 spherical regions of radius 22 (comoving) Mpc were then resimulated using a full hydrodynamics prescription \citep{Rasia2015} with the {\tt GADGET-X} SPH solver \citep{Beck2016}. Haloes and subhaloes were identified with Amiga's Halo Finder \citep{Knollmann2009}, which has a binding energy criterion similar to {\sc Subfind}, but uses an adaptive mesh refinement grid to represent the density field/contours. 

While $\The$ is only mass-complete above $\Mtwohc \gtrsim 10^{15}$~$\msol$ at $z=0$, it resolves many haloes below this mass. Since the scaling relations derived from these lower mass clusters are in agreement with our other simulations, we note that the selection effects do not in general appear to have an impact on our temperature scaling results, with one exception. That is, within large radii (i.e., $\Rvir$ and $\Rtwohm$) the scatter of temperatures is amplified, however, the population mean behaviors remain unaffected.

In Appendix~\ref{sec:Res_Feedback}, we also use two different simulations from $\The$, which we will refer to as {\sc Music} and {\sc Gizmo}. {\sc Music} uses the solver {\tt GADGET-MUSIC} \citep{Sembolini2013} in place of {\tt GADGET-X}. A summary of the differences between these simulations can be found in \cite{Cui2018}, but in brief, the {\tt GADGET-MUSIC} and {\tt GADGET-X} solvers use similar but subtly different gas treatments and stellar formation and stellar feedback mechanisms. Most notably, however, {\tt GADGET-MUSIC} does not include any Black Hole or AGN feedback, while {\tt GADGET-X} does.

{\sc Gizmo} \citep{Cui2022} on the other hand uses {\tt GIZMO-SIMBA} \citep{Dave2019} in its Meshless Finite Mass solver mode instead of {\tt GADGET-X} which uses SPH. A detailed discussion of the differences can be found in \citet[their Table 2]{Cui2022}. In general, {\sc Gizmo} uses a different feedback model, with, among other things, significantly stronger kinetic feedback calibrated for high mass haloes which causes large differences in the observed gas fractions between the {\tt GIZMO-SIMBA} and {\tt GADGET-X} runs.

\vspace{-4mm}
\subsection{Magneticum Pathfinder}
{\sc Magneticum Pathfinder} \citep{Hirschmann2014} is a suite of magneto-hydrodynamics simulations using a version of the SPH solver {\tt GADGET-3} developed independently to that used in the $\BaM$ versions. We used the {\tt Box2 hr} run, and will henceforth refer to it as the $\Mag$ sample. Haloes are once again identified using a FoF algorithm, and subhaloes using {\sc Subfind}.
We note that since the $\Mag$ simulations are based on WMAP7 cosmology \citep{Komatsu2011} they have the most distinct cosmology to the other simulations which all use Planck cosmologies \citep{Planck2014XVI, Planck2016XIII}.
Finally, we observe that while the other simulations show little variation between the core-excised and non-core excised values for $\Tm$ and $\Ty$ \citep[as discussed for $\BaM$ in][]{Lee2020}, this is not true in $\Mag$. As such, we use the core-excised values for all three temperature measures obtained from $\Mag$. This is explored in more detail in Appendix~\ref{app:core-excision}.

\vspace{-4mm}
\subsection{IllustrisTNG}
The {\sc IllustrisTNG} project, here $\Tng$, \citep{Springel2018, Pillepich2018FirstResults, Nelson2018, Marinacci2018, Naiman2018} is a follow up to the {\sc Illustris} project \citep{Vogelsberger2014}. It uses the moving mesh code {\tt AREPO} \citep{Springel2010} and uses a full magneto-hydrodynamics treatment with galaxy formation models as detailed in \citet{Pillepich2018SimulatingGalaxyFormation, Weinberger2017}. We use TNG300-1, the highest resolution run from the suite, but reference the two lower resolution runs in Appendix~\ref{sec:Res_Resolution}. Haloes are identified using a FoF algorithm and subhaloes with {\sc Subfind}, as for the $\Mag$ sample.

We estimate all TNG properties using the FoF particle set. The FoF linking length of $b = 0.2$ was chosen so that the FoF group on average contains all particles within $\Rtwohc$ of the halo center. Consequently, properties computed within significantly larger apertures (primarily $\Rtwohm$) will miss the contribution from particles in the far outskirts (as these are not included in the FoF, whose particle set is incomplete at such distances) and will thus incur a minor bias. However, our main analysis and results focus on $\Rtwohc$ and $\Rfivehc$ and are thus unaffected by this bias.

\begin{figure}
    \includegraphics[width=\linewidth]{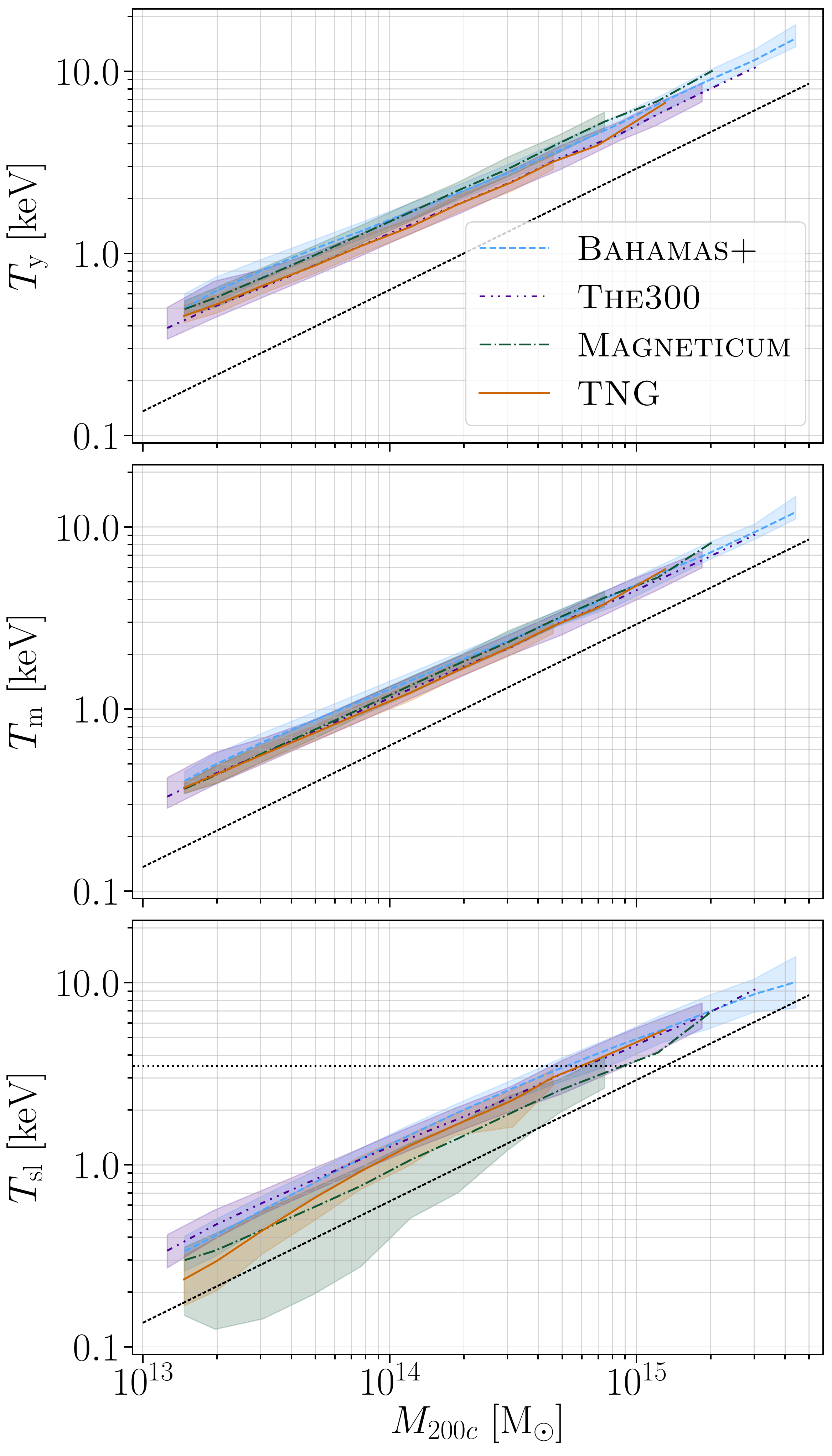}\\
    \caption{The three temperature measures in each simulation at $z=0$. Here the data is binned into logarithmically spaced mass bins. The shaded regions show the 16 and 84 percentile regions within these bins, while the lines show the medians. The dotted black line indicates the self-similar scaling $T\propto M^{2/3}$, whose amplitude is arbitrarily set for visual clarity. The horizontal line in the lower panel indicates the temperature above which $\Tsl$ is considered a good proxy for the X-ray temperature.}
    \label{fig:03_TvM_z0}
\end{figure}
\vspace{-4mm}
\section{Temperature Scaling Relations}
\label{sec:TS}
In this section, we briefly discuss the temperature measures at redshift $z=0$, before examining how they vary with redshift (Section~\ref{sec:TS_Redshift}) and the choice of the radial cut-off (Section~\ref{sec:TS_Radial}). We also examine how the temperatures scale with respect to the Compton-$y$ parameter, and the $\Ty$ offset against $\Tm$ itself, in Sections~\ref{sec:TS_Y} and \ref{sec:TS_DeltaT}. In Section~\ref{sec:Res}, we discuss how the specifics of the simulations themselves have affected our results. 

\begin{figure*}
    \includegraphics[width=0.8\linewidth]{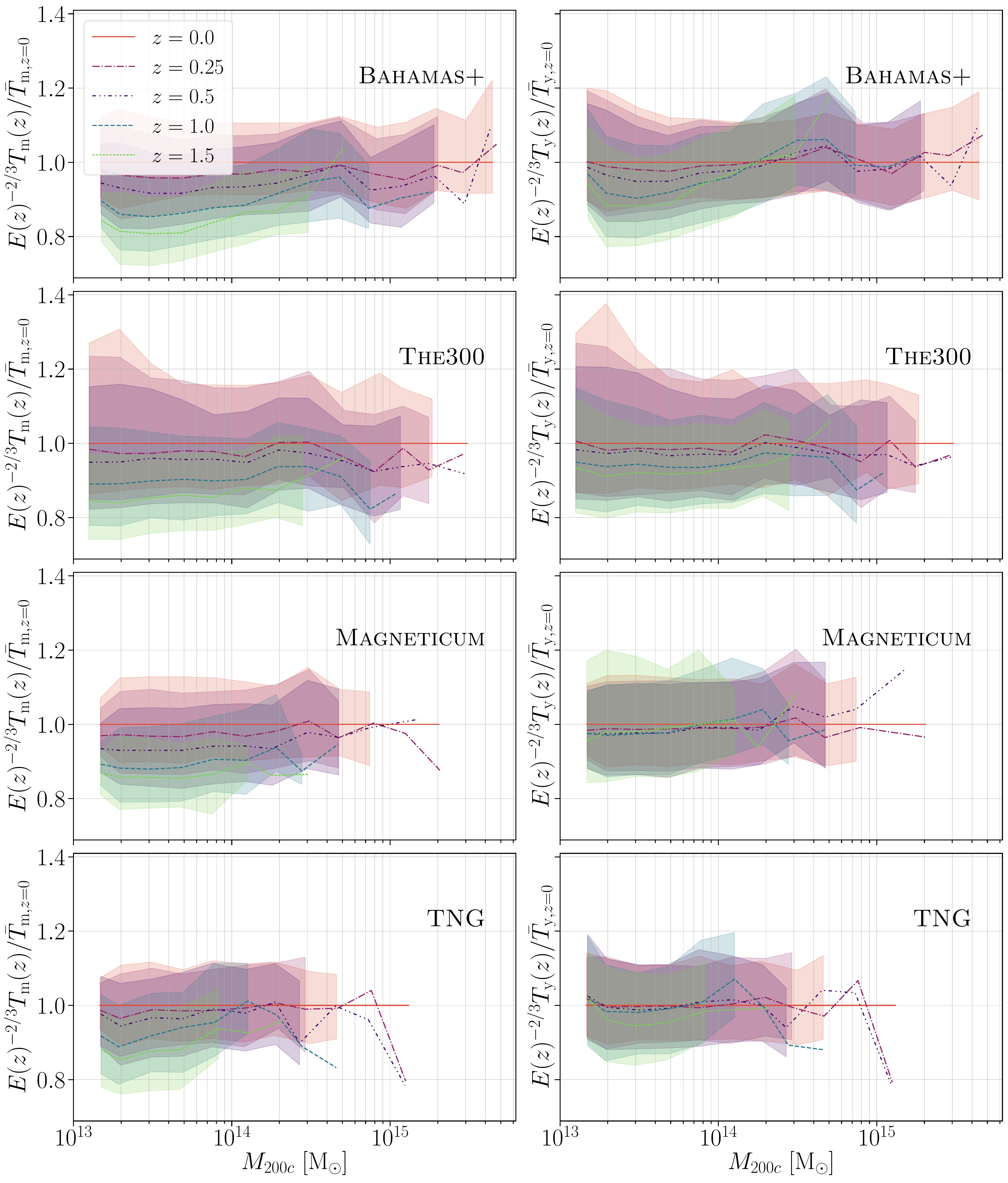}\\
    \caption{Redshift evolution of the temperature measures within each simulation. As in Fig.~\ref{fig:03_TvM_z0}, the data is sorted into mass bins, with the shaded regions showing the 16 to 84 percentile region and the lines the medians. Here each redshift for each sample is divided by the bin median at $z=0$. We see there is reduced evolution in $\Ty$ (the right panels) than in $\Tm$ (those on the left).}
    \label{fig:04_TvM_redshift}
\end{figure*}
In general, we can see a good agreement between the 4 samples for each temperature measure (Fig.~\ref{fig:03_TvM_z0}). As we will continue to use a similar plotting convention throughout this paper, we briefly explain the process here. We have sorted the data into logarithmically spaced mass bins, and within each bin then plotted the median, and 16\% and 84\% percentile regions for each sample. That is, the solid lines indicate the median of the data distribution, while the shaded regions show the 1$\sigma$ intrinsic scatter within each data set. Where there are fewer than 10 clusters within a mass bin, we have only plotted the median and not calculated the percentiles. Within each sample, this region accounts for an intrinsic variation of around $\pm 5.5\%$ for $\Tm$ and $\Ty$ (with this, in general, being marginally larger in $\Ty$ than $\Tm$), and $\pm 9\%$ for $\Tsl$.

We can see that for the mass-weighted temperature, $\Tm$, all four simulations show a close agreement at $z=0$, while $\Ty$ shows slightly more variation, but is, nonetheless, overall consistent. The spectroscopic-like temperature shows the most variation between the four samples, particularly at lower masses/temperatures. However, it is worth noting, that for $\Tsl \gtrsim 3.5$~keV, where $\Tsl$ is considered a good proxy for the X-ray temperature, the samples agree well.
In general, however, there is far more intrinsic scatter within the $\Tsl$ measures (within $\Rtwohc$) than for the other temperature measures. We can see in particular that $\Mag$ has a large scatter in $\Tsl$, especially at lower masses. This is driven by the warm gas ($T<10^6$~K) in the low mass haloes, and is discussed in more detail in Appendix~\ref{app:T-cut}. Due to our limited $\Tsl$ data where it is an appropriate proxy for the X-ray temperature, particularly at higher redshifts, we will not consider $\Tsl$ in detail for the rest of this section.

In Fig.~\ref{fig:03_TvM_z0}, we have also plotted an arbitrary indicative line to show self-similar scaling, i.e., $T\propto M^{2/3}$. This allows us to see that in simulations all three temperature measures appear to scale at slightly less than $2/3$, with $\Tsl$ lying closest to this. However, at higher masses and temperatures, $\Ty$ and $\Tm$ appear to tend to this self-similarity, while at lower masses and temperatures, the scaling relation seems shallower. Conversely, this also indicates there is some mild curvature within the $\Ty$ and $\Tm$ scaling relations. The two and three parameter fits for each of these simulations can be found tabulated in Table~\ref{apptab:fits_median_z0_samples} for a more detailed comparison, and will be discussed further in Section~\ref{sec:cross_sim_fits}.
The agreement at high masses may come from the decreased relative effects of feedback in this regime. That is, at lower masses the gas is hotter than expected from solely gravitational heating due to feedback processes, pushing the equilibrium away from self-similarity, while at higher masses the potential well ensures more gas is retained in haloes; this is supported by the results of \citet{Farahi2018}, who use kernel-localized linear regression \citep{Farahi2022} to show that the gas mass in \textsc{Bahamas+} clusters approaches a self-similar scaling at higher halo masses. That the lower masses lie higher than expected from self similarity for $\Ty$ and $\Tm$, would indicate that generally the feedback is leading to hotter gas in the halo.

\subsection{Redshift evolution}
\label{sec:TS_Redshift}
\begin{table}
\caption{The median value for $E(z)^{-2/3}T(z)/\bar{T}_{z=0}$ as shown in Fig.~\ref{fig:04_TvM_redshift}, for $z=0.5$ and 1.0. These can be understood as a broad mass-independent redshift correction. However, we have tabulated in fact the median correction to the 2-parameter fit for each simulation (described in Section~\ref{sec:Dis_effects} and tabulated themselves in Table~\ref{apptab:fits_median_z0_samples} in Appendix~\ref{app:tables}).} \centering
\begin{tabular}{lccc}
 $z=0.5$ &  $E(z)^{-2/3}\Ty$ &  $E(z)^{-2/3}\Tm$ & $E(z)^{-2/3}\Tsl$ \\ 
 \hline \smallskip
 $\BaM$ & $0.981^{+0.002}_{-0.002}$ & $0.951^{+0.001}_{-0.001}$ & $0.929^{+0.005}_{-0.011}$ \\ \smallskip
 $\The$  & $0.968^{+0.006}_{-0.004}$ & $0.953^{+0.003}_{-0.002}$ & $0.933^{+0.007}_{-0.005}$ \\ \smallskip
 $\Mag$  & $0.969^{+0.002}_{-0.005}$ & $0.932^{+0.003}_{-0.003}$ & $0.728^{+0.023}_{-0.014}$ \\ \smallskip
 $\Tng$  & $0.984^{+0.007}_{-0.005}$ & $0.972^{+0.004}_{-0.005}$ & $0.746^{+0.018}_{-0.038}$ \\
 \hline
 $z=1.0$ &  $E(z)^{-2/3}\Ty$ &  $E(z)^{-2/3}\Tm$ & $E(z)^{-2/3}\Tsl$ \\ 
 \hline \smallskip
 $\BaM$ & $0.955^{+0.003}_{-0.002}$ & $0.896^{+0.002}_{-0.002}$ & $0.815^{+0.006}_{-0.011}$ \\ \smallskip
 $\The$  & $0.938^{+0.004}_{-0.004}$ & $0.903^{+0.002}_{-0.002}$ & $0.856^{+0.004}_{-0.005}$ \\ \smallskip
 $\Mag$  & $0.966^{+0.004}_{-0.003}$ & $0.889^{+0.003}_{-0.005}$ & $0.590^{+0.020}_{-0.013}$ \\ \smallskip
 $\Tng$  & $0.997^{+0.007}_{-0.007}$ & $0.941^{+0.006}_{-0.005}$ & $0.501^{+0.018}_{-0.012}$ \\
 \hline
\end{tabular}
\label{tab:fits_redshift_offset_samples}
\end{table}
As discussed in Sect.~\ref{sec:MvR_theory}, from self-similarity we expect the cluster temperatures to scale as $T(z) \propto E(z)^{2/3}$ at fixed mass. In Fig.~\ref{fig:04_TvM_redshift}, we can see that this is not quite the case within our sample.
We examine our temperature measures at 5 different redshifts, $z=0.0$, 0.25, 0.5, 1.0 and 1.5. In general, we find that the temperature measures increase (with increasing redshift) {\it slower} than self-similarity would suggest. That is, in Fig.~\ref{fig:04_TvM_redshift}, were self-similar evolution to occur, we would expect all 5 redshifts to align -- as is nearly true for $\Ty$. We note that here we have used the cosmological parameters and true redshift for each individual simulation to calculate this redshift behaviour. Due to the larger differences in cosmology in the $\Mag$ simulation, this allows for more consistent results between simulations.

We found that $\Tsl$, while not plotted here, evolves with the least accordance to self-similarity, while $\Ty$ diverges the least from it. That is, graphically, we can see the spread in the 5 redshift means is {\it smaller} for $\Ty$ than for $\Tm$. At $z=1$, we can see that $\Tm$ has a median lowered to around the 1$\sigma$ intrinsic scatter at $z=0$. $\Ty$ however shows almost no redshift dependence in $\Mag$ and $\Tng$ and only mild evolution in $\BaM$ and $\The$.
Physically, we can motivate this aspect, as at higher redshifts, clusters on average had shorter cooling times, leading to more cool dense gas, which down-weights $\Tsl$. On the other hand, $\Ty$ is determined by the gas pressure, which, assuming hydrostatic equilibrium, would be fixed to match the size of the potential well itself, leading to a more self-similar temperature measure.

It is also interesting to note that in general our samples do not show overt mass dependence in the redshift evolution for the two SZ temperatures, i.e., graphically, the mean lines are roughly horizontal for all of our samples. Here, it is worth noting that we are considering the masses at the redshift the cluster temperature is measured. In Table~\ref{tab:fits_redshift_offset_samples}, we quantify these offsets for all three temperature measures in each sample at $z=0.5$ and $1.0$. Here we can more quantitatively see that there is increasing divergence from self-similarity (value of unity), comparing $\Ty$ to $\Tm$ to $\Tsl$ in all cases. We can also see the variety in redshift behaviour for each sample. We also note that the larger errors in the $\Tsl$ offsets arise from there being more mass dependence on the redshift evolution (particularly for $\Mag$ and $\Tng$) than in $\Ty$ and $\Tm$. The full details of the $\Tsl$ are not considered in much detail as the haloes included at these higher redshifts in $\Mag$ and $\Tng$ are rarely at temperatures high enough for $\Tsl$ to be an accurate prediction of the observed X-ray temperature.

\subsection{Radial dependence}
\label{sec:TS_Radial}
\begin{figure*}
    \includegraphics[width=0.8\linewidth]{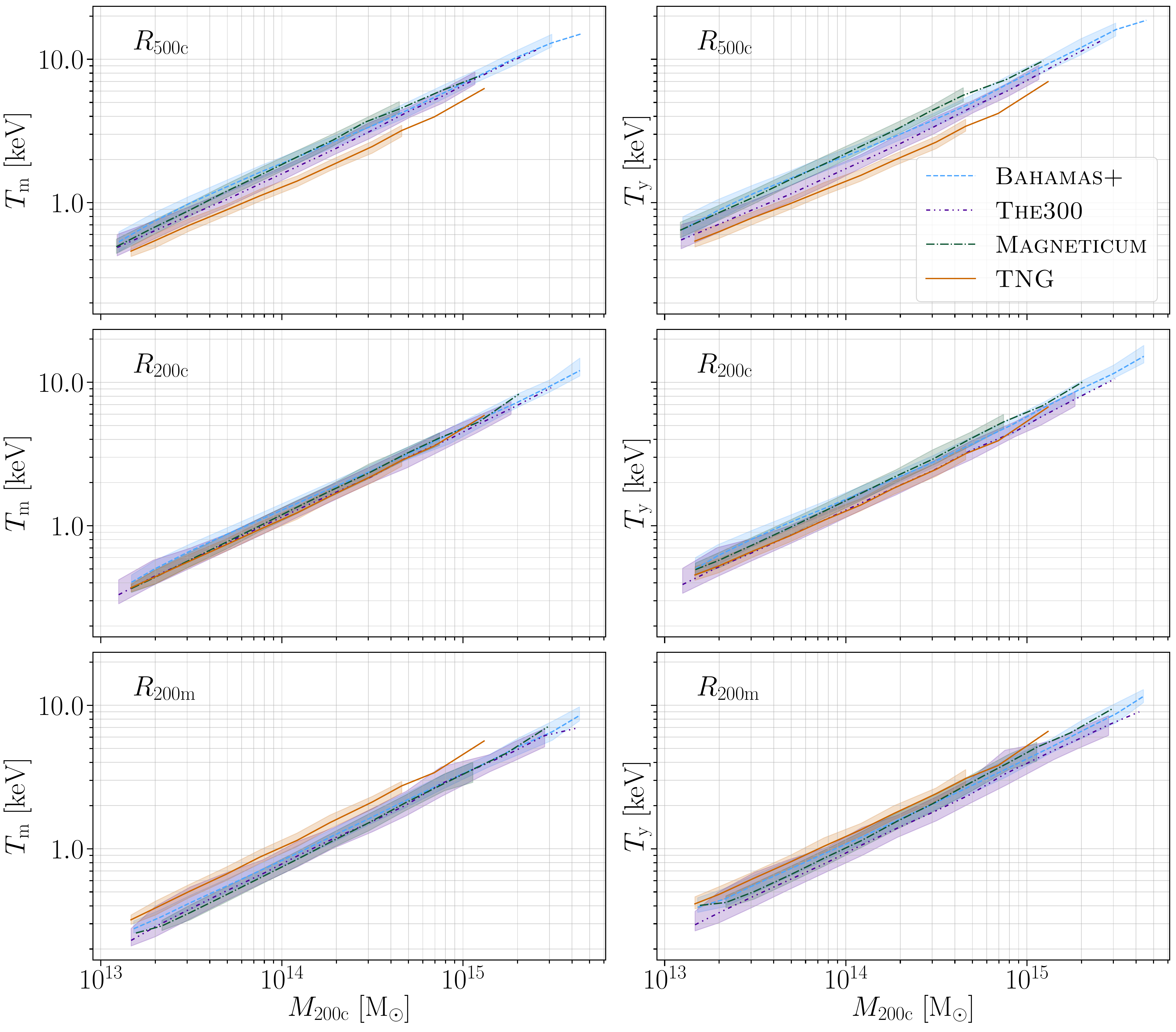}\\
    \caption{The variation in temperature scalings when averaged over three different radii, at $z=0$. This figure is arranged as in Fig.~\ref{fig:03_TvM_z0}, with $\Tm$ on the left and $\Ty$ on the right.}
    \label{fig:06_radial_scalings}
\end{figure*}
\begin{figure}
    \includegraphics[width=\linewidth]{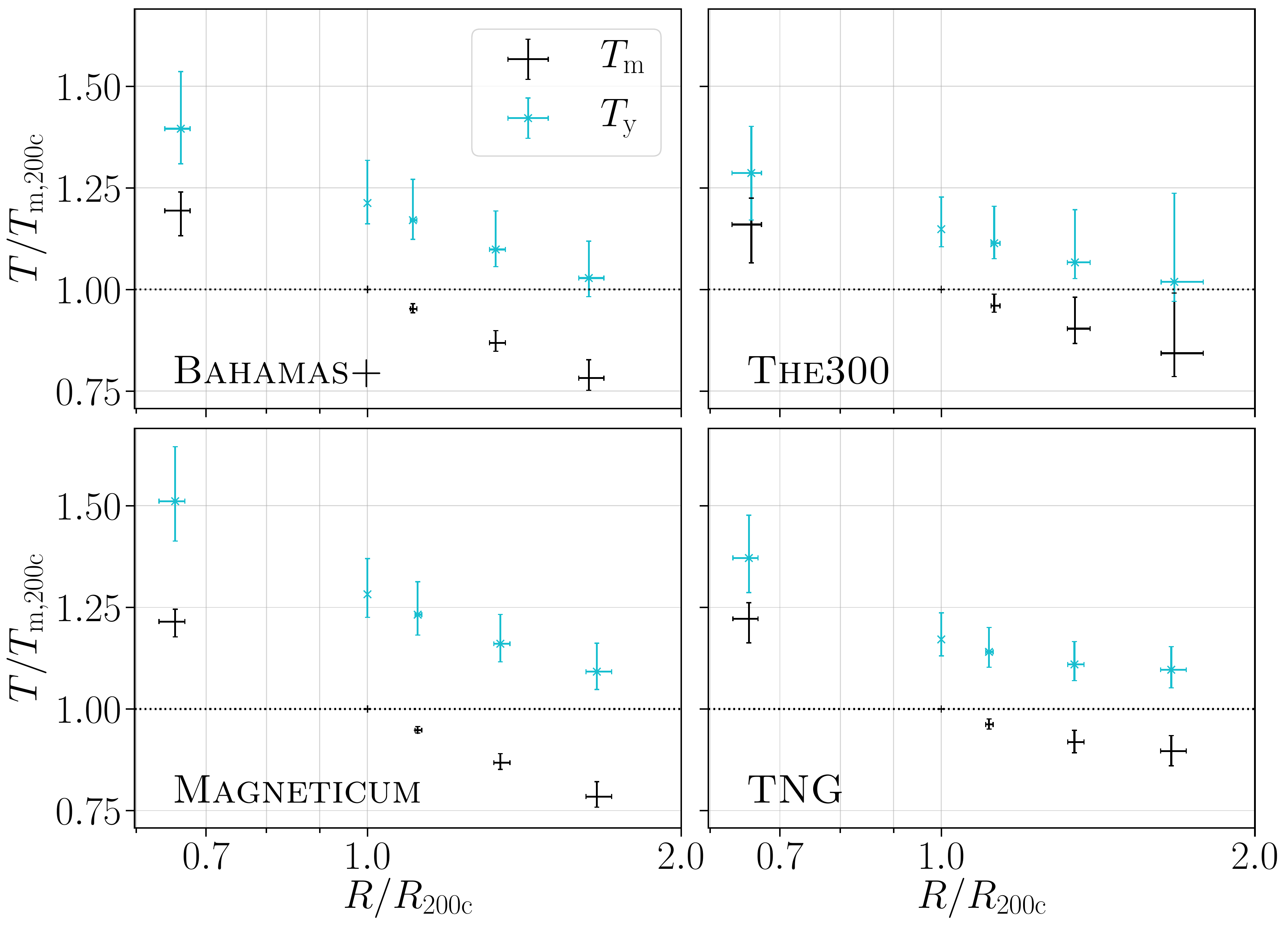}\\
    \caption{A depiction of the radial dependence of each simulation. Here every cluster's temperature and radius is divided by the {\it same cluster's} radius and $\Tm$ within $\Rtwohc$, for the five radii considered. These are, from left to right $\Rfivehc$, $\Rtwohc$, $\Rfivehm$, $\Rvir$, and $\Rtwohm$. Hence the error bars show the intrinsic scatter within each simulation for the clusters' temperature profiles, fixed at $\Rtwohc$ and $T_{\rm m,200c}$.}
    \label{fig:05_radial_scatter}
\end{figure}

Next, we study the dependence of the temperature measures on the radius of the sphere that we average over. Fig.~\ref{fig:05_radial_scatter} shows how the averaged temperatures vary over the five radii we consider at $z=0$. Here, on a cluster-by-cluster basis, we divide each radius and temperature of the cluster, by the $\Rtwohc$ and $T_{\rm m,200c}$ values for that cluster, and then we show the bulk averages of these values. These values can be found for each sample for convenient reference in Table~\ref{apptab:R_err_sample} in the Appendix.

We find that the profiles are very similar within the four samples, and are consistent out to $\Rfivehm$. The variations beyond this are likely driven by the particulars of the simulations rather than reflecting any inconsistency. We discuss in Section~\ref{sec:Simulations} that when calculating our $\Tng$ values, only particles that are linked to the FoF group are included, which may bias the large radius temperatures high. $\The$ shows a consistent profile with the other simulations, but a larger scatter due to the sample selection for the low mass haloes. \citet{Anbajagane2022} found similar scatter amplification in the velocity dispersion of low mass clusters in \textsc{The300}, and explicitly showed this was generated by the selection effect (see Figure B1).
We can also see that the offset between $\Ty$ and $\Tm$ is larger in $\BaM$ and $\Mag$ than in the other two simulations -- this will be explored more in Section~\ref{sec:TS_DeltaT}. This offset also seems to marginally increase at larger radii.

The difference in profile steepness can be appreciated in Fig.~\ref{fig:06_radial_scalings}. Here we show the temperature-mass relations, at $\Rfivehc$, $\Rtwohc$, and $\Rtwohm$, so we can see how the variation in profiles depends on the masses of the clusters. For example, we find that the differences in the profiles between $\The$ and $\BaM$ are largely driven by lower mass haloes, and agree well at all radii for higher masses.

It also becomes evident that, for both $\Tm$ and $\Ty$, the four models agree best within the $\Rtwohc$. At lower radii, the shallow profile of $\Tng$ corresponds to temperatures below those in the other simulations, and at larger radii to temperatures above. Similarly, the steeper profile of $\Mag$ leads to its temperatures sinking with respect to the other samples, as the radius increases. $\Tsl$, while not shown here, has a larger variation at each radius -- and it is harder to determine the agreement between simulations due to the difference in intrinsic scatter in each sample. Within $\Rfivehc$, $\Tsl$ in $\Mag$ agrees far better with the other simulations -- however, $\Tsl$ in all the other simulations agree slightly worse with each other than at $\Rtwohc$. 

\subsection{Temperature--Y scalings}
\label{sec:TS_Y}
\begin{figure}
    \includegraphics[width=\linewidth]{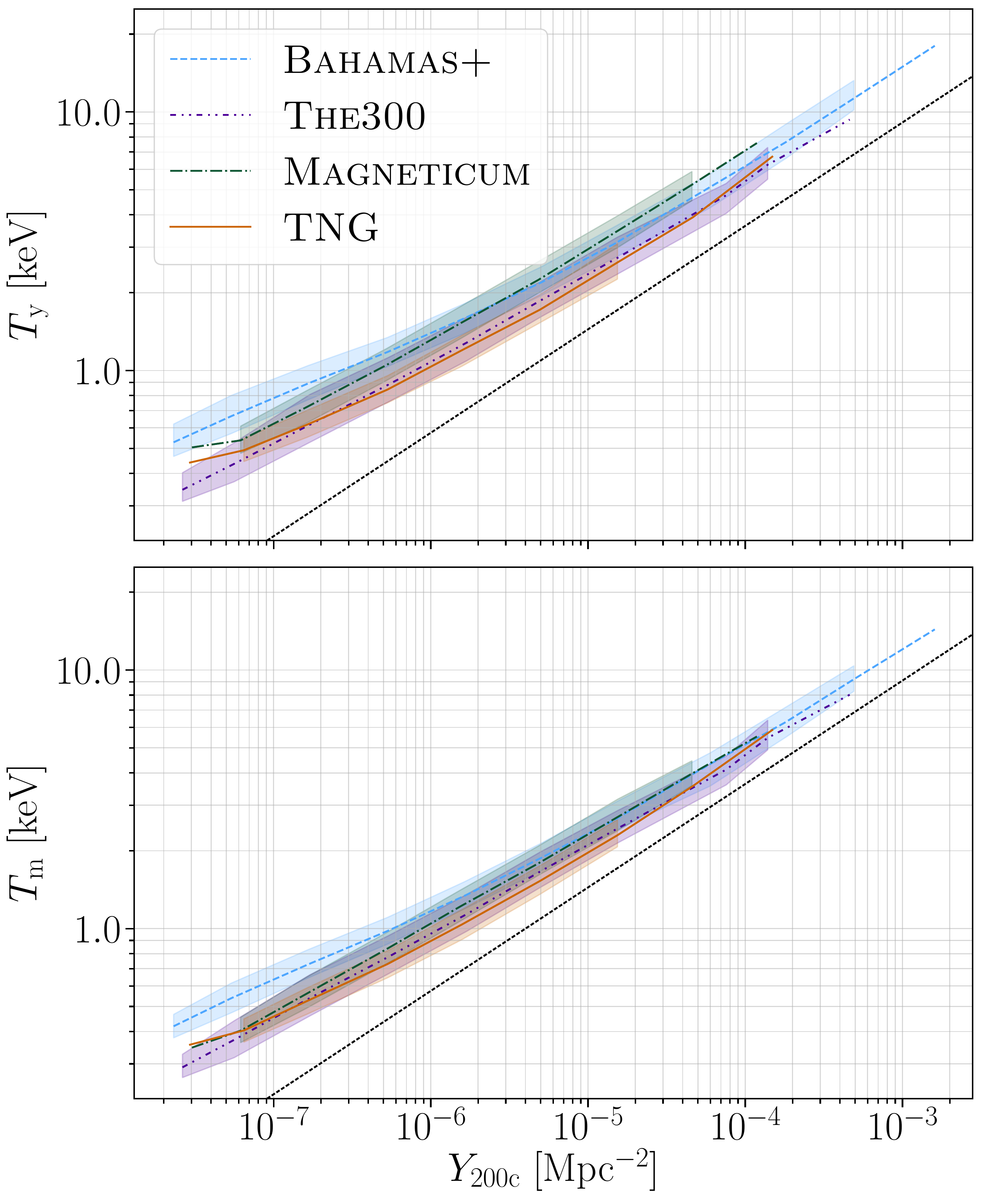}\\
    \caption{The variation of each temperature measure with respect to the volume-averaged Compton-$y$ within $\Rtwohc$. These are the measures at $z=0$ and the figure is arranged as in Fig.~\ref{fig:03_TvM_z0}, grouped into logarithmically spaced Y bins, rather than mass bins. The dotted black line is an indicative line with scaling $T\propto Y^{2/5}$, i.e., a self-similar scaling.}
    \label{fig:07_TvY_scaling}
\end{figure}
While temperature-mass scaling relations are important and relevant for comparison with X-ray observations, a consideration of the Compton-$y$ parameter in relation to temperature could lead to a way to self-calibrate SZ observables \citep{Lee2020}. In Fig.~\ref{fig:07_TvY_scaling}, we show how the SZ temperatures scale within our samples with $Y_{\rm 200c}$.
We can immediately see that there is slightly poorer agreement within these quantities than when using $\Mtwohc$, however, these do still predominantly agree between the different samples. In particular, as $Y_{\rm 200c}$ increases, the agreement between $\Tm$ in each sample improves, and as we will see later in Section~\ref{sec:Res}, most of this variation correlates with the variation of $\fgas$ in each simulation. 

We note that the $Y_{\rm 200c}-\Mtwohc$ relationship agrees very well between simulations for all masses $\gtrsim 10^{14}$~$\msol$, which corresponds to $Y_{\rm 200c} \gtrsim 10^{-6}$~Mpc$^{-2}$. It is also worth considering that since our haloes are selected with a mass cut-off, for the lowest values of $Y$ ($\lesssim 10^{-7}$~Mpc$^{-2}$) the data we have are not necessarily complete for each value of $Y$, so may be biased slightly high.
Self-similarity would suggest a scaling relation of $T \propto Y^{2/5}$, but we see shallower scaling relations $\simeq 0.33$ in all our samples. In particular, $\BaM$ has a shallower scaling relation ($\simeq 0.31$), for both $\Ty$ and $\Tm$ than the other three samples. 

\subsection{$\Delta T$--Temperarature scalings}
\label{sec:TS_DeltaT}
\begin{figure}
    \includegraphics[width=\linewidth]{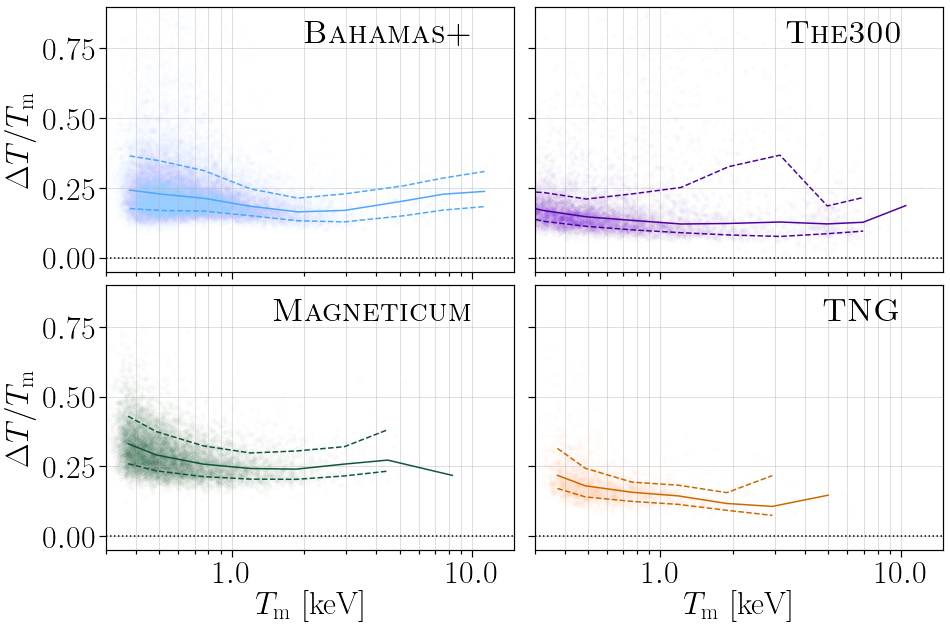}\\
    \caption{The relative variation in temperature measures, $\Delta T= \Ty-\Tm$ with respect to the mass-weighted temperature, $\Tm$. Here each point represents a single halo within the simulation, the solid line represents the medians within logarithmically spaced $\Tm$ bins, while the dashed lines show the 16 and 84\% percentiles for the same bins. The discrepancy in the scatter for $\The$ around $T\simeq 3$~keV is discussed below in Section~\ref{sec:TS_DeltaT}.}
    \label{fig:09_DTvT_scaling}
\end{figure}
As already mentioned, we find an offset between the temperature measures, which we now explore in more detail. For this, in Fig.~\ref{fig:09_DTvT_scaling}, we plot the fractional temperature difference, $\Delta T/\Tm = (\Ty-\Tm)/\Tm$ against $\Tm$ itself. We already identified that the mass-weighted temperature is a good proxy for the mass itself, so it is worth noting that this temperature difference has a similar, if subtly different, relationship than when plotted against mass.

We can see immediately that the systematic offset is different in each of the four samples. However, it is interesting to note that this offset is subject to a significant skew -- which is to say, at the simplest level the offset holds on a cluster by cluster basis, and $\Ty$ is {\it always} greater than $\Tm$ within clusters.
This can be seen visually, as here we have plotted every halo within our samples in Fig.~\ref{fig:09_DTvT_scaling}, and it is evident that within each cluster, there is a minimum difference, greater than zero, between all the clusters. Taking the 1st percentile for each cluster sample we can estimate this minimum offset as being highest in $\Mag$ with $\Delta T \geq 0.178\, \Tm$, and smallest in $\The$ where $\Delta T \geq 0.066\, \Tm$.

Here, it is worth briefly discussing why the $\The$ sample has such a non-uniform intrinsic scatter compared to the other samples. We note that this divergence is worst at $\Tm \simeq 3$~keV, reaching convergence again at high temperatures, ($\Tm \gtrsim 4$~keV). This is an artifact of the selection process for low mass clusters within the $\The$ sample. That is, since all `low mass' ($\Mtwohc\lesssim 0.9\times10^{15}$~$\msol$) clusters, exist in the region of larger haloes, this biases the temperatures of the clusters. As such, this region is less relevant for a mass-complete understanding of the temperature differences here.

\subsection{Resilience of results}
\label{sec:Res}
\begin{figure}
    \includegraphics[width=\linewidth]{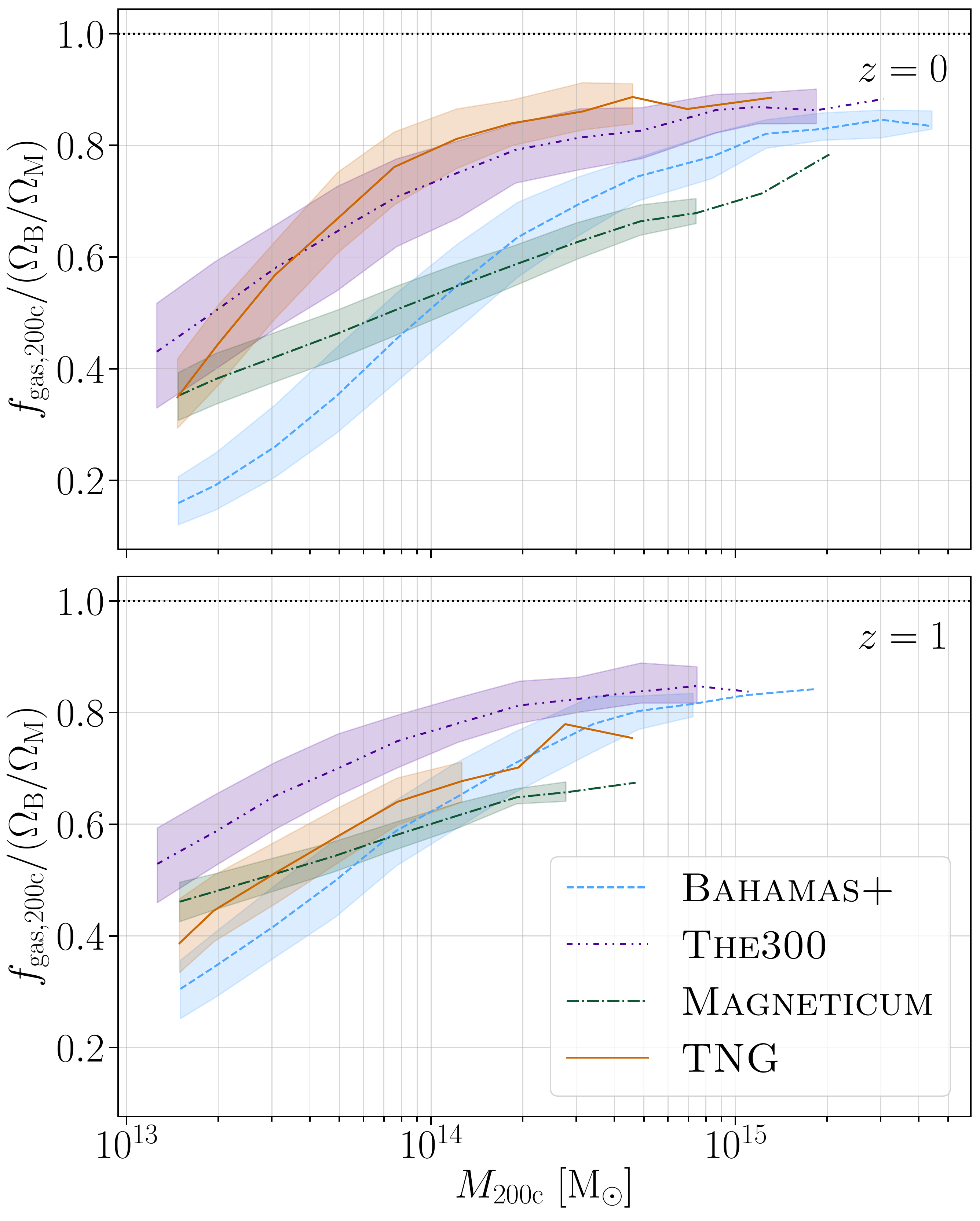}\\
    \caption{A comparison of the gas fraction between each simulation. Here the gas fraction, $f_{\rm gas, 200c}=M_{\rm gas,200c}/M_{\rm 200c}$ has been normalised by the relevant $\Omega_{\rm B}$ and $\Omega_{\rm M}$ for each simulation. This figure is arranged as in Fig.~\ref{fig:03_TvM_z0}, with $z=0$ in the top panel and $z=1$ in the lower panel.}
    \label{fig:10_GasFraction_vM}
\end{figure}
All the simulations use different physical models and numerical methods to generate the halo populations we study. Moreover, they are all calibrated to different measurements, as such it is remarkable that we see the agreement we have found across the four different samples. However, in this section we will focus more specifically on the effects these differences have caused on our observed results.

Firstly, we briefly consider the effect of resolution in simulation -- this is presented in more detail in Appendix~\ref{sec:Res_Resolution}. We have found inconclusive results: We used three different resolution runs from $\Tng$ where all the other parameters were held the same, and found minimal effects on $\Tm$ and $\Ty$. $\Tsl$ did decrease in value with increasing resolution, most likely due to its dependence on resolving small dense clumps in clusters. However, we also compared two different resolution runs $\Mag$, where each was independently calibrated (although these results are not plotted in this work). Here the lower resolution run also resulted in lower values for $\Tm$ and $\Ty$, with a shift around that of the intrinsic variance within each simulation. Within the context of \citet{Schaye2015}, this indicates that the temperatures show `strong' resolution convergence, but show a diminished `weak' convergence.

Since our temperature measures are dependent on gas density, it is important to consider how the differences in gas fraction (i.e., $f_{\rm gas, 200c} = M_{\rm gas, 200c}/M_{\rm 200c}$.) vary between our simulations. In Fig.~\ref{fig:10_GasFraction_vM}, we show the gas fractions within $\Rtwohc$ (scaled by the cosmic baryonic fraction in the simulations) at $z=0$ and 1. In general, we often consider $\fgas$ to be a probe of the feedback used in clusters, with lower values for $\fgas$ indicating more effects from feedback. Hence generally speaking, at lower masses, $\fgas$ is lower as the potential well for haloes is shallower, allowing for more gas to be ejected from haloes from feedback.

It is evident that there are significant variations in $\fgas$ among the simulations considered in this work. At the highest masses, $\Tng$, $\The$, and $\BaM$ start to agree, with values $>0.9$, but $\Mag$ lies greatly below this. Similarly, at the lowest masses, at $z=0$, $\Mag$, $\The$ and $\Tng$ roughly agree, while $\BaM$ is significantly smaller. 

Furthermore, the samples exhibit a varying level of redshift evolution. The gas fraction rises generally, indicating that as clusters evolve, generally feedback results in gas being ejected from haloes, so younger clusters (i.e., in general, clusters at higher redshifts), will have higher values for $\fgas$. However, the slopes in each of the samples decrease by differing values -- $\Tng$ most notably, with its values reducing dramatically everywhere except the lowest mass haloes.

This diverging behaviour in $f_{\rm gas, 200c}$, lies in contrast to the strong agreement we find between the samples for both the $\Tm-$mass relations, and the $Y-$mass relations, although it is indicated by the $T-Y$ variation. Since $Y$ is a measure of the gas pressure within clusters, we could expect it to be self-calibrating -- that is, a cluster in hydrostatic equilibrium, would lead to a certain gas pressure to counter the gravitational well of its own mass. As such we would expect haloes to have a strong $Y-M$ relationship.
 
However, we can see a reflection of the $\fgas$ variation subtly in $\Ty$. That is, we can see that the curvature of the $\Ty-$mass relation as seen in, e.g., Fig.~\ref{fig:03_TvM_z0}, is similar (albeit inversely) to the curvature in the $\fgas-$mass relation. In fact, we can find that this is driven by the changes in $\Mgas$ as $\Mgas \Ty$ forms a tight relation with $Y$ and a slightly more consistent relation (than $\Ty$ alone) with mass. This is perhaps unsurprising, as we already know that on a cluster by cluster level, $\Ty$ and $\Tm$ form a strong relationship, and since $\Mgas \Tm \propto Y$, we can consider $\Mgas \Ty$ as a kind of relativistic equivalent to $Y$. Nonetheless, this means a more detailed understanding of $\Mgas$ (or equivalently $\fgas$) in clusters would lead to more assurance in the exact $\Ty-$mass and $\Ty-Y$ relationships.

In Appendix~\ref{sec:Res_Feedback}, we compare three different feedback model runs used in the $\The$ project. In general, we find that $\Tm$ is barely affected by the different models, only perturbed in regions where the feedback models are extreme. In contrast, $\Ty$ matches our suppositions from studying $\fgas$, where again $\Ty$ is largely consistent across the different feedback models, but shows variations inversely related to the variation of $\fgas$. That is to say, stronger feedback in general leads to higher values of $\Ty$, but variations in the specific implementation of feedback lead to more complicated effects. $\Tsl$ varies significantly in amplitude and gradient between the three feedback runs, making it a potential probe for the `true' feedback in clusters, but less reliable as a temperature proxy.

\section{Cross simulation averaged results}
\label{sec:cross_sim_fits}
Due to the broad agreement of all of the samples in each temperature, we can consider the effects of averaging across simulations to obtain `simulation-independent' predictions for these values from simulations. Here we obtain these by joining all the samples to form one large population of halos, which we then sort into mass bins as before, and find sample fits to the mass-binned averages.\footnote{This approach obtains results consistent with those obtained by, for instance, taking the mass-binned values for each sample, and joining these together.} We note that this means our averages will be weighted more by those simulations with larger populations of halos. We also provide fits to individual simulations in Table~\ref{apptab:fits_median_z0_samples}, and the variations in simulation predictions can be easily used to estimate the theoretical, astrophysics-driven uncertainty in the mean temperature-mass relations, for example via the method described in \citet[see Section 4.3]{Anbajagane2022}.

When fitting this data to obtain temperature-mass relations, we then consider both a two- and three-parameter fit of the form
\begin{equation} \label{eqn:fit_format}
    T = E(z)^{2/3} A \left(\frac{M}{10^{14}\;\msol}\right)^{B+C\log_{10}(M/10^{14}\msol)}\,{\rm keV},
\end{equation}
where for the two-parameter fit we set $C=0$. The cross-simulation averaged results at $z=0$ for all three temperatures can be found in Table~\ref{tab:fits_median_z0_all} -- the fits for each individual sample at $z=0$ can be found in Table~\ref{apptab:fits_median_z0_samples}. The errors are obtained through bootstrap techniques and show the error within the mean. Here we have also calculated a measure of the scatter through the root mean squared dispersion around these mean values as
\begin{equation} \label{eqn:sigma_def}
    \slog=\sqrt{\frac{1}{N}\sum_i\left[\log_{10}\left(\frac{T_i}{T_{\rm fit}(M_i)}\right)\right]^2},
\end{equation}
with $i$ indexing over all the halos at a given redshift. We note that this measure is weighted more by the lower mass clusters and groups, due to the larger number of them in each sample than higher mass haloes. This runs opposite to our fits where we have minimised this bias by fitting to the averages gained from a selection of mass bins. 

\begin{table}
\caption{The two and three parameter fits [Eq.~\eqref{eqn:fit_format}] and a measure of the intrinsic scatter [Eq.~\eqref{eqn:sigma_def}] in the fits against mass within $\Rtwohc$ at $z=0$ for the cross simulation averaged sample. The errors here are the errors in the fits found through bootstrapping.} \centering
\begin{tabular}{lcccc}\smallskip
 & $A$ & $B$ & $C$ &$\slog$ \\ 
 \hline \smallskip
 $\Ty$  & $1.465^{+0.002}_{-0.002}$ & $0.586^{+0.003}_{-0.002}$ & 0 & $0.1025^{+0.0003}_{-0.0002}$\\ \smallskip
 $\Ty$  & $1.426^{+0.006}_{-0.007}$ & $0.566^{+0.001}_{-0.001}$ & $0.024^{+0.005}_{-0.004}$ & $0.1011^{+0.0001}_{-0.0001}$\\ 
 \hline \smallskip
 $\Tm$  & $1.210^{+0.001}_{-0.001}$ & $0.591^{+0.003}_{-0.003}$ & 0 & $0.0805^{+0.0002}_{-0.0001}$\\ \smallskip
 $\Tm$  & $1.207^{+0.005}_{-0.005}$ & $0.589^{+0.001}_{-0.001}$ & $0.003^{+0.004}_{-0.005}$ & $0.0804^{+0.0001}_{-0.0001}$\\
 \hline \smallskip
 $\Tsl$ & $1.135^{+0.003}_{-0.003}$ & $0.601^{+0.006}_{-0.011}$ & 0 & $0.2067^{+0.0016}_{-0.0008}$ \\ \smallskip
 $\Tsl$ & $1.196^{+0.020}_{-0.009}$ & $0.641^{+0.003}_{-0.003}$ & $-0.048^{+0.007}_{-0.018}$ & $0.2028^{+0.0002}_{-0.0001}$\\
 \hline
\end{tabular}
\label{tab:fits_median_z0_all}
\end{table}

The first aspect we can note is the exceptional lack of curvature in the $\Tm-$ Mass relationship, where even when we allow for curvature, $C$, is fully consistent with zero, albeit with a slight tendency toward positive curvature. This is true even if we vary the pivot mass of the fit. We also see that $\Ty$ has definitive positive curvature, as follows from our early discussion. On the other hand, $\Tsl$ shows some significant negative curvature, but this may be an artifact of the lower mass haloes' scatter, and may not be representative of the behaviour of observed X-ray temperatures.

It is also important to note that the intrinsic variance is slightly larger in the combined sample than it is in each simulation, due to small bulk offsets between each sample. It should be reiterated, however, that the errors tabulated in Table~\ref{tab:fits_median_z0_all} show the errors in the fit parameters, obtained through bootstrapping -- as such these do not reflect the 16 and 84 percentile lines plotted in our figures. They also are not only caused by the variation in the mean due to the bulk offsets between simulations. At $z=0$, the intrinsic variance can be found to be around $\pm6.0\%$ in $\Tm$, $\pm7\%$ for $\Ty$ and around $\pm10\%$ in $\Tsl$. However, a more detailed understanding of the intrinsic variance can be found by looking at the cross-simulation fits to the 16 and 84 percentile lines which can be found in Tables~\ref{apptab:fits_median_z0_all} and \ref{apptab:fits_3_median_z0_all}.

In general, it is also important to note that there is still a fixed offset between $\Ty$ and $\Tm$ to be found in the combined sample -- we find a median offset of around 22\% (see Table~\ref{tab:R_err_all}). However, as with the discussion in Section~\ref{sec:TS_DeltaT}, we can consider the minimal offset to be determined by the 1\% percentile line -- in this case, giving a minimum offset of $8.4\%$. This large variance in $\Ty/\Tm$ is of course, largely driven by the slight disagreement between clusters, which could potentially be broken with future measurements of $\fgas$.

We also use this cross-simulation sample to calculate the redshift evolution of these quantities, and the radial dependence of these results. The first order redshift and radial evolution can be seen in Tables~\ref{tab:fits_redshift_offset_all} and \ref{tab:R_err_all} akin to those discussed in Section~\ref{sec:TS}. The mean 2-parameter fits can be found in Table~\ref{apptab:fits_median_Redshift_all}, alongside the full set of 2- and 3-parameter fits in Tables~\ref{apptab:fits_R5_median_z0_all} and \ref{apptab:fits_3_R5_median_z0_all} for the three temperatures within $\Rfivehc$ at $z=0$.

\begin{table}
\caption{The median correction to the amplitude, $A$, alone of the two-parameter fit at each redshift for the temperature measures in the cross-simulation averaged simulation. That is, we are tabulating the averaged value $A^\star$, so that $E(z)^{-2/3} T_z = A^\star T_{z=0}$ The full two-parameter fits for each redshift can be found in Table~\ref{apptab:fits_median_Redshift_all}.} \centering
\begin{tabular}{lccc}
 &  $E(z)^{-2/3}\Ty$ &  $E(z)^{-2/3}\Tm$ & $E(z)^{-2/3}\Tsl$ \\ 
 \hline \smallskip
 $z=0.00$ & 1.000 & 1.000 & 1.000 \\ \smallskip
 $z=0.25$ & $0.982^{+0.002}_{-0.002}$ & $0.979^{+0.002}_{-0.001}$ & $0.960^{+0.017}_{-0.011}$ \\ \smallskip
 $z=0.50$ & $0.976^{+0.002}_{-0.002}$ & $0.952^{+0.001}_{-0.001}$ & $0.929^{+0.004}_{-0.007}$ \\ \smallskip
 $z=1.00$ & $0.944^{+0.002}_{-0.002}$ & $0.897^{+0.002}_{-0.002}$ & $0.837^{+0.007}_{-0.011}$ \\ \smallskip
 $z=1.50$ & $0.919^{+0.002}_{-0.003}$ & $0.851^{+0.002}_{-0.002}$ & $0.749^{+0.004}_{-0.004}$ \\
 \hline
\end{tabular}
\label{tab:fits_redshift_offset_all}
\end{table}
\begin{table}
\caption{The shifts and errors in the radius, mass, $\Tm$, and $\Ty$ over the combined sample against those values within $\Rtwohc$. These values are calculated on a cluster-by-cluster basis, and then the averages are found within these. The central value here is the median with the errors given by the 16 and 84 percentiles.} 
\centering
\begin{tabular}{lcccc}\smallskip
 & $R/\Rtwohc$ & $M/\Mtwohc$ & $\Tm/T_{\rm m,200c}$ & $\Ty/T_{\rm m,200c}$ \\ 
 \hline \smallskip
 $\Rfivehc$  & $0.66^{+0.01}_{-0.02}$ & $0.71^{+0.05}_{-0.07}$ & $1.20^{+0.04}_{-0.07}$ & $1.40^{+0.16}_{-0.12}$\\  \smallskip 
 $\Rtwohc$  & $1.00$ & $1.00$ & $1.00$ & $1.22^{+0.11}_{-0.07}$\\  \smallskip 
 $\Rfivehm$  & $1.11^{+0.01}_{-0.01}$ & $1.08^{+0.03}_{-0.02}$ & $0.95^{+0.01}_{-0.01}$ & $1.18^{+0.10}_{-0.06}$\\  \smallskip 
 $\Rvir$ & $1.33^{+0.03}_{-0.02}$ & $1.22^{+0.08}_{-0.05}$ & $0.87^{+0.04}_{-0.02}$ & $1.11^{+0.09}_{-0.06}$\\  \smallskip 
 $\Rtwohm$ & $1.64^{+0.06}_{-0.04}$ & $1.39^{+0.15}_{-0.10}$ & $0.79^{+0.07}_{-0.04}$ & $1.05^{+0.09}_{-0.06}$\\
 \hline
\end{tabular}
\label{tab:R_err_all}
\end{table}

We can see in Table~\ref{tab:fits_redshift_offset_all} that the sample averaged redshifts show the same variation we expected from considering each sample independently. That is, all of the temperatures diverge from self-similarity, with $\Tsl$ showing the greatest departure, and $\Ty$ the smallest. However, when looking at high redshift haloes, some departure is nonetheless to be expected. Again, we reiterate that the errors here are the errors in the fit, and as such do not encapsulate any variation in the intrinsic scatter from combining samples. It is also interesting to note that, although not shown here, the $\slog$ values for both these 1-parameter corrections and the 2-parameter fits that are tabulated in \ref{apptab:fits_median_Redshift_all} are broadly the same. Although, as previously noted, these will be largely weighted by the lower mass haloes, this does indicate that most of the evolution is in the amplitude of the scaling relations, and not in their power law.

This redshift evolution is also not linear with respect to redshift, changing faster at lower redshifts. However, we can find a parameterisation of this redshift evolution in the form of 
\begin{equation}
    \log_{10}(A^\star) = p \log_{10}(1+z)+ q \log^2_{10}(1+z),
\end{equation}
where $A^\star$ is the redshift correction factor, so $T_z = A^\star E(z)^{2/3} T_{z=0}$. We find for $\Ty$, $\Tm$ and $\Tsl$ that the values for $[p,q]$ are $[-0.05,-0.11]$, $[-0.08,-0.24]$ and $[-0.09, -0.57]$, respectively. However, it is worth noting that at higher redshifts, $\Tsl$ becomes an increasingly poor proxy for $T_{\rm X}$ as the number of haloes we have at temperatures above 3.5~keV diminishes. 

The radial corrections tabulated in Table~\ref{tab:R_err_all} {\it show} the effects of combining the samples, where the errors represent the intrinsic variance. We can see that the combined sample only slightly increases this intrinsic radial variance. While the effects of changing the radius are more complex than a single number can fully encapsulate, this is still useful for an indicative understanding of the effects on viewing clusters through different apertures. However, it is interesting to note that the value of $\Ty/\Tm$ at each radius, increases at higher radii. That is, as we increase the radius of interest, $\Tm$ decreases faster than $\Ty$ or equivalently, $\Ty$ has a shallower profile. The equivalent values for each sample to those in Table~\ref{tab:R_err_all}, can be found in Table~\ref{apptab:R_err_sample}.

\begin{figure}
    \includegraphics[width=\linewidth]{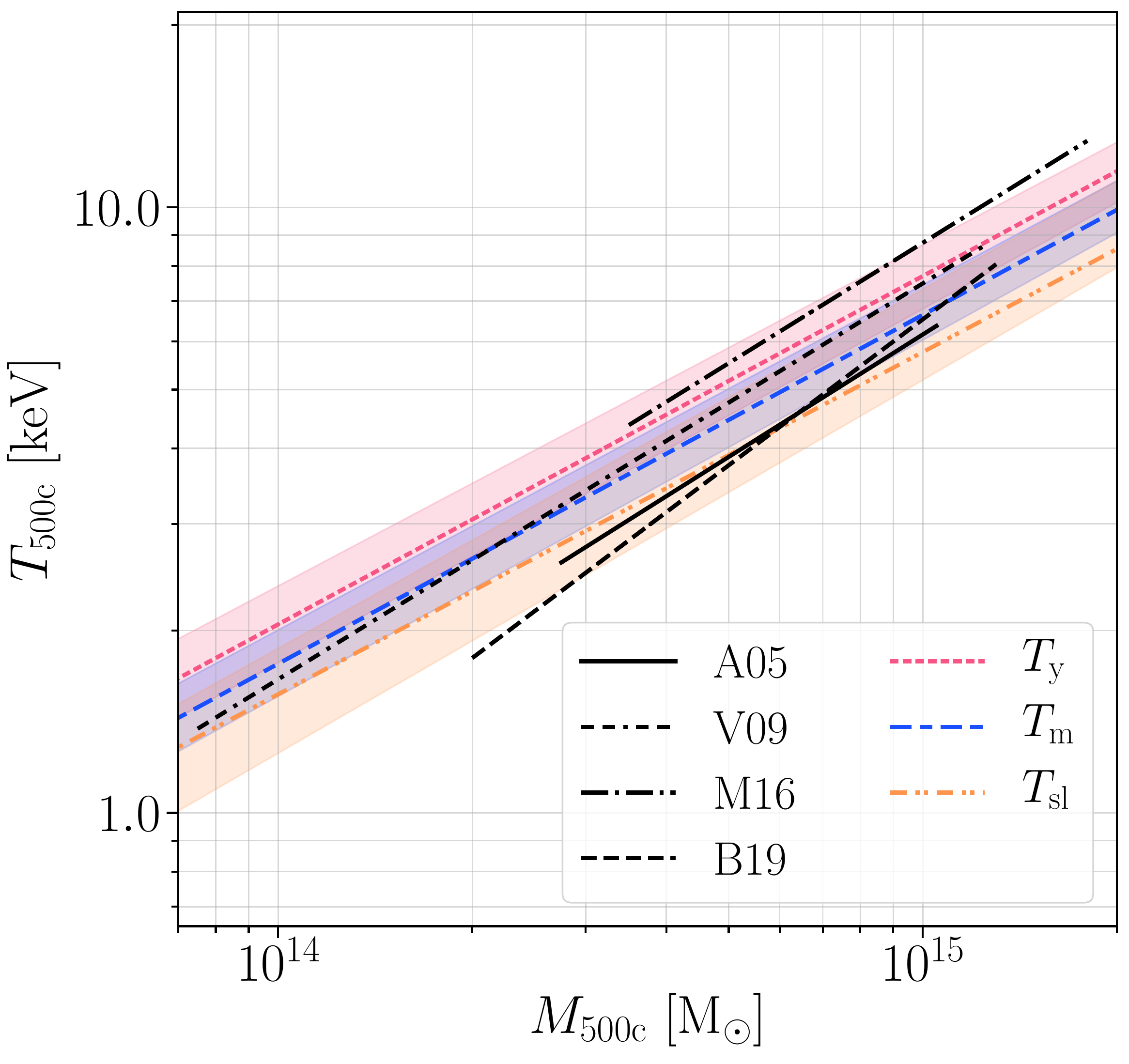}\\
    \caption{A comparison of the cross simulation-averaged temperature scaling relations, with observational results. Note these quantities are all measured within $\Rfivehc$. Here the observational lines are plotted only within the mass range of each study. Here A05, V09, M16 and B19 refer to the fits from \citet{Arnaud2005,Vikhlinin2009xrayobs,Mantz2016} and \citet{Bulbul2019}, respectively. A05 and V09 both depend on the hydrostatic mass bias, for which we here assumed $1-b=0.8$.}
    \label{fig:15_observations}
\end{figure}
\subsection{Comparison to X-ray observations}
There have been many observational studies aiming to constrain the X-ray temperature, $T_{\rm X}$, mass relationship. Here, we will consider 4 such studies to compare to our cross-simulation averaged results. In particular, we have considered \citet[A05]{Arnaud2005}, \citet[V09]{Vikhlinin2009xrayobs}, \citet[M16]{Mantz2016} and \citet[B19]{Bulbul2019}, all of which provide 2 parameter fits for $T_{\rm X}$-$\Mfivehc$ relation. 

A05 uses 10 low redshift ($z<0.15$) clusters from {\it XMM-Newton}, and obtains masses through fitting an NFW-type profile assuming hydrostatic equilibrium. Here we use their fits for the subsample of 6 hot ($>3.5$~keV) clusters.
V09 uses {\it Chandra} measurements of 85 clusters (over redshifts of $z\lesssim0.9$), obtaining through fits of $\beta$-profiles, again assuming hydrostatic equilibrium. As such, both A05 and V09's masses are offset from the `true' mass according to the hydrostatic mass bias. This has been measured many times, both in simulations and using observational techniques, and in this work, we will use $(1-b)=0.8$.
M16 uses 40 hot ($>5$~keV), relaxed clusters observed by {\it Chandra} (with redshifts $z\lesssim 1$). This work calibrates its masses using weak lensing measurements, so can be taken to be measures of the true mass.
Finally, B19 uses 59 clusters from {\it XMM-Newton}, over redshifts $0.2<z<1.5$. The masses are obtained through SZ measurements by the South Pole Telescope. This work also provides two different fits for $T_{\rm X}$ and here we use their core-excised fits, to compare best with our other measures.

In Fig.~\ref{fig:15_observations}, we show these 4 fits against our whole sample fits, here all within $\Rfivehc$. The length of each line matches the mass range of the data sets used within each observation. It is first worth noticing, that there is no strong consensus between X-ray observations about the details of the $T_{\rm X}$-$M$ relationship. This may be exacerbated by the different techniques used in each study. For B19 and M16 an intrinsic scatter of around 13\% is given, while V09 obtains a slightly higher 15-20\%. These are calculated as $\sigma_{{\rm ln}T}$ and lie a little smaller than those values predicted from our combined simulations.

Here A05 and B19 both use data sets from {\it XMM-Newton} and agree with each other best, as well as holding comparable results to the whole-sample $\Tsl$ measurement. However, B19 has a far steeper gradient (at $B=0.80^{+0.11}_{-0.08})$ than any of the other studies, which is a huge contrast to all of our simulation fits which indicates gradients are always less than self-similarity (that is $B<2/3$). The other three studies, also all have gradients steeper than that obtained through our simulations, all of which are consistent with self-similarity. It is interesting to note, that the full sample for A05, (rather than the hot sample displayed here) also shows a shallower gradient more compatible with our simulation results.

M16 shows the most extreme result, tending towards far higher temperatures than seen in the other scalings, and higher even than the $\Ty$ values, we have found within simulations. This may be a side effect of the sample selection towards hot clusters, pushing the average up, or may indicate the intrinsic spread within the $T_{\rm X}$ relationship. V09, the other measurement using {\it Chandra} observations, also lies higher than the {\it XMM-Newton} values, indicating an X-ray temperature more consistent with our simulation $\Tm$ values, than $\Tsl$.

It is still unclear whether the variation observed here between the observational methods and simulations, indicates that simulations are not capturing some facet of real galaxy clusters, or are miscalibrated, or if this comes from the difficulties in obtaining `true' masses from observations. A factor that could be exacerbated by the suggested intrinsic spread in the $T_{\rm X}$ relationship and the comparatively (when compared to the data sets from simulations) small data sets and mass ranges used within observational studies.
Either way, the $\Ty$ temperature is not directly accessible by any of these X-ray studies and we thus instead recommend using the scalings derived here in future modeling of SZ observables. 

\section{Discussion} 
\label{sec:applications} 
\subsection{$\Ty$--$Y$ self-calibration}
\label{sec:self-calibration}

The $Y$--$\Ty$ relation can be used in the SZ signal modeling once the $Y$-parameter is determined. This allows one to refine the SZ model (with the relativistic SZ corrections) even if the data for individual systems cannot provide a direct constraint on $\Ty$. The observed Y can also differ from the true Y due to angular resolution (caused by the instrument beam) but when the resolution is $\lesssim$ 1 arcmin, these smoothing effects in $Y$ are negligible \citep[][]{Yang2022}, and so the true $Y$--$\Ty$ relation presented here can be used directly on observed data without needing any further processing. Similarly, one can easily use the $Y$--$\Ty$ relation in simulations of the SZ sky, where from the simulation the cluster temperature is not directly available, e.g., in the {\tt WebSky} \citep{Stein2020}.

Here we will briefly discuss the scaling relationship we have determined, alongside the effects we may be able to determine with these results. As noted, we find some complexity in this relationship, with varying levels of curvature, especially at small values of $Y$. As such, we have tabulated the 2- and 3-parameter fits for the $\Tm$-$Y$ and $\Ty$-$Y$ relationships in Table~\ref{apptab:fits_Y_percentiles_z0_all}. The equivalents for the medians for each sample can also be found in Table~\ref{apptab:fits_Y_median_z0_sample}.

\begin{table}
\caption{The two-parameter $\Ty$-$Y$ fits [Eq.~\eqref{eqn:Yfit_format}] for haloes with $Y_{\rm 200c}>10^{-6}$~Mpc, and a measure of the intrinsic scatter [Eq.~\eqref{eqn:sigma_def}] in the fits across redshifts for the median of the cross simulation averaged sample within $\Rtwohc$.} \centering
\begin{tabular}{lccc}
 & $A$ & $B$ & $\slog$ \\ 
 \hline \smallskip
 $z=0.00$ & $2.614^{+0.006}_{-0.006}$ & $0.368^{+0.003}_{-0.010}$ & $0.1874^{+0.0039}_{-0.0152}$ \\ \smallskip
 $z=0.25$ & $2.593^{+0.007}_{-0.007}$ & $0.372^{+0.002}_{-0.005}$ & $0.1685^{+0.0036}_{-0.0074}$ \\ \smallskip
 $z=0.50$ & $2.593^{+0.008}_{-0.008}$ & $0.381^{+0.003}_{-0.002}$ & $0.1625^{+0.0034}_{-0.0033}$ \\ \smallskip
 $z=1.00$ & $2.585^{+0.013}_{-0.012}$ & $0.382^{+0.004}_{-0.004}$ & $0.1388^{+0.0044}_{-0.0045}$ \\ \smallskip
 $z=1.50$ & $2.597^{+0.052}_{-0.031}$ & $0.373^{+0.020}_{-0.011}$ & $0.1263^{+0.0185}_{-0.0091}$ \\
 \hline
\end{tabular}
\label{tab:fits_Yhot_median_redshift_all}
\end{table}

We can, however, create more stable 2-parameter fits if we exclude the smallest halos with $Y<10^{-6}$, which are unlikely to meaningfully contribute to SZ observations, and which, even when they can be observed, will have the smallest rSZ corrections. With this restriction and 
\begin{equation} \label{eqn:Yfit_format}
    T = E(z)^{2/5} A \left(\frac{Y}{10^{-5}\;\msol}\right)^{B}\,{\rm keV},
\end{equation}
we can find fits for the combined sample at each redshift as given in Table~\ref{tab:fits_Yhot_median_redshift_all} (a full form including the 16 and 84 percentiles can be found in Table~\ref{apptab:fits_Yhot_percentiles_redshift_all}). We can immediately see that there is little redshift dependence beyond the expected self-similar evolution.

We note that even here we have more intrinsic variation in this relationship than within the $\Ty-M$ relationships; however, it can still be used as a reliable proxy to estimate the relativistic effects to haloes. Moreover, as our understanding of $\fgas$ in clusters improves, we may be able to reduce this uncertainty between clusters, as this would allow a greater understanding of the true $\Mgas$ values we should expect for haloes, and thus a more precise idea of exactly how we expect the $\Ty-Y$ relationship to behave.

With those comments in mind, we can immediately gain a sense of the corrections we would expect for a given cluster. For instance, from our simulations, we can estimate that a cluster with a true $Y_{\rm 200c} = 10^{-4}$ at $z=0$, would have a temperature of $6.1^{+0.8}_{-1.1}$~keV, where the errors here are driven almost entirely by the disagreement between the simulations. This would lead to a fractional change in the amplitude at 353~GHz of $0.907^{+0.016}_{-0.011}$ -- that is, a 10\% underestimation of $Y$ by using the non-relativistic approximation.

It is also important to remember that the relativistic corrections to the kinetic SZ (kSZ) effect, will be driven by $\Tm$ and not $\Ty$. We have not discussed the kSZ effect in much detail in this work, but nonetheless, detailed studies of kSZ signals, should also consider the impact that this signal may have. These relativistic kSZ effects, are, however, broadly speaking an order of magnitude smaller than the relativistic tSZ correction \citep[e.g.,][]{Sazonov1998, Nozawa2006, Chluba2012}. Still, the relations given here can be used to model the effects.

\subsection{Applications to current and future SZ analyses}
\label{sec:Dis_effects}
In this section, we briefly discuss where the obtained scaling relations may have applications in current and future studies of the SZ effect. First and foremost, due to a lack of sensitivity or spectral coverage, current SZ measurements are still not directly sensitive to the rSZ effect. Nevertheless, rSZ can already affect the inference of cosmological parameters if it is neglected in the modeling. One important example is related to the SZ power spectrum analysis, for which rSZ can lead to an underestimating of the $yy$-power spectrum and thus cause a systematic shift in the inferred value of $\sigma_8$ \citep{Remazeilles2019}. They would similarly impact cross-correlations of the SZ field with large-scale structure fields such as galaxy positions or cosmic shear, and such measurements have been recently used to infer physics like the redshift-dependent mean thermal pressure of the Universe as well as the energetics of feedback in groups and clusters \citep[e.g.,][]{Osato2018, Osato2020, Pandey2019, Pandey2021, Gatti2021}.

The size of the rSZ bias depends directly on the experimental configuration. Similarly, different experimental setups are more or less prone to mis-modeling of rSZ. For example, many of the planned CMB experiments have channels at $\nu \lesssim 220\,{\rm GHz}$. In this case, the degeneracy of the SZ signal with respect to the $y$-parameter and $\Ty$ cannot be easily broken in the observation. However, our new scaling relations will allow us to accurately include the rSZ corrections in the theoretical modeling of the data.

Similar to SZ power-spectrum analyses, the SZ cluster number count analyses are expected to be affected by the presence of rSZ. Here, two aspects are important: at a given mass, the SZ flux is diminished due to rSZ. This means that clusters are assigned to a lower signal-to-noise bin. In addition, for the extraction of the clusters, the multi-match-filtering (MMF) method \citep{Melin2006} is not optimally tuned to the correct spectral shape, leading to a misestimation of the noise. The scaling relations introduced here can be directly used to inform the MMF and eliminate the impact rSZ may have. In particular, the $\Ty-Y$ relation will be of significance here, since it allows to construct an iterative MMF approach \citep[e.g.,][]{Zubeldia2022} to incorporate the rSZ effect based solely on SZ observables. This should allow for a more robust comparison of theory and observation and also improve the constraining power of the obtained SZ catalogs.

The rSZ also impacts studies on the thermodynamics of cluster outskirts ($R \gtrsim \Rtwohm$) --- especially non-thermal features like cosmological accretion shocks \citep{Aung2021, Baxter2021ShocksSZ} --- which contain astrophysical and cosmological information \citep[see][for a review]{Walker2019} and have only recently been observationally explored. \citet{Anbajagane2022c} performed the first large population-level analysis of tSZ profile outskirts and found signs of cosmological shocks, which manifest as decrements in the profile. Others have seen similar signs using samples of up to ten clusters \citep[e.g.,][]{Hurier2019ShocksSZPlanck, Pratt2021ShocksPlanck}. However, as was described above, the rSZ effect \textit{also} causes a tSZ signal reduction. Thus, a better understanding of the rSZ, and the magnitude of the tSZ decrement it causes, will be needed to robustly infer the non-thermal \citep{Nelson2014b,Aung2021} and plasma physics in these outskirts \citep{Rudd2009,Avestruz2015}. 

As we have seen in Fig.~\ref{fig:04_TvM_redshift}, $\Ty$ has a scaling relation, close to self-similar evolution with redshift. However, nonetheless, significant redshift evolution is still expected from self-similarity, and the minor corrections we observe. This effect can have important consequences for cosmological inferences relying on the redshift-independence of the SZ effect. One example is attempts to use the SZ effect to measure the CMB temperature-redshift relation \citep{Rephaeli1980, Luzzi2009}, another relates to SZ-derived values of the Hubble constant \citep{Birkinshaw1991, Mauskopf2000, Wan2021}. Similarly, applications of SZ measurements to constrain possible time-variations of the fine structure constant \citep{Bora2021} are prone to the redshift-dependence of rSZ effect. With the derived $\Ty-Y$ relations, the effect of rSZ on these inferences can be readily marginalized over. 

As a final application of rSZ we mention predictions of the all-sky averaged SZ and rSZ effects \citep{Hill2015}. This signal is one of the targets of future CMB spectroscopy \citep{Chluba2019} and can inform us about feedback processes in cluster physics \citep[e.g.,][]{Thiele2022}. The temperature relations obtained here allow us to refine these predictions, giving a simulation-averaged view on the expected signal. In a similar manner, the relations can be used to refine the calculations of the radio \citep{Holder2022, Lee2022} and CIB SZ \citep{Sabyr2022, Acharya2022} effects. These signals might become important targets for future radio and sub-mm observations, allowing us to probe the evolution and origin of the cosmic radio and infrared backgrounds.

\section{Conclusions}
\label{sec:Conclusion}
In this work, we present detailed comparisons of three cluster temperature measures: (a) the average rSZ temperature, (b) the mass-weighted temperature relevant for the thermal SZ (tSZ) effect, and (c) X-ray spectroscopic temperature using the $\Bah$ \& $\Mac$, {\sc Illustris-TNG}, $\Mag$, and {\sc The Three Hundred Project} simulations. We analyze gas temperature scaling relations of galaxy groups and clusters with $\Mfivehc>10^{13}$~$\msol$, over 5 redshifts between $z=0$ to $z=1.5$. We provide fits to multiple scaling relations for individual simulations and for a combined cross-simulation sample, with the former ensemble of results also providing an estimate of the theoretical, astrophysics-driven uncertainty in the relations.
Our main results are summarized as follows:
\begin{itemize}
    \item There is an exceptionally strong agreement for $\Tm$ between all four simulations with $\Mtwohc$. $\Ty$ is consistently larger than $\Tm$ (by an average of $\simeq22\%$), which is generally a little above $\Tsl$ at $z=0$. $\Ty$ has a good agreement between the simulations, however, there is variation in the exact magnitude of the offset between $\Ty$ and $\Tm$ between simulations. $\Tsl$ also shows agreement, although it is also subject to a great deal more intrinsic scatter than the other two temperatures. All three temperature measures exhibit different mass scalings and vary differently with redshift and radius.
    \item All three temperature measures exhibit deviation from the self-similar evolution. At higher redshifts, they all fall below the expected $E(z)^{-2/3}$ scaling indicating all haloes will have lower temperatures at higher redshifts than the self-similar model prediction. However, $\Ty$ evolves very closely to self-similarity, while the other temperature measures depart further from self-similarity, so that at higher redshifts, all three temperature measures increasingly diverge. $\Ty$ has an increasing correction relative to $\Tm$ and an even larger correction to $\Tsl$.
    \item The temperature measures all agree best within $\Rtwohc$ between simulations. Each simulation and temperature measure has a different radial profile leading to varied results. While the temperature measures still agree well within $\Rfivehc$, the improvement at $\Rtwohc$ indicates this may indeed be an optimal radius to study SZ science.
    \item The gas fraction, feedback methods, and resolution all vary significantly between the simulations. In light of this, the level of agreement we see is startling, and indicates that much of the SZ gas physics is sufficiently calibrated by the microphysical constraints (i.e., stellar properties). However, when we examine these in more detail, we can find that the gas fraction is correlated with the variation in $\Ty$ between simulations. As such, if this can be determined with more accuracy in future observations, the strength of these predictions may increase. In general, when we study resolution and feedback within equivalently calibrated simulations, we find little variation in our observed SZ temperature measures, while $\Tsl$ is affected slightly more.
    \item We have created a cross-simulation sample and found the fitted values for our temperatures. In general, $\Tm$ shows a limited tendency towards curvature (i.e., a mass-dependent slope), while $\Ty$ has positive curvature, and $\Tsl$ has negative curvature. We have provided a simple regime for calculating the redshift corrections to temperatures within our redshift range and clarified the broad effects of varying the radial aperture which we use to define haloes. When compared to observational results, we find that the temperatures broadly agree. However, there is more variation within X-ray results than our predictions, making it difficult to draw out strong conclusions. Nonetheless, observations all suggest steeper scaling relations than we have found in the simulations.
    \item In general while these temperatures will be difficult to directly measure, they give rise to the possibility of self-calibrating SZ observations, to allow for relativistic corrections to be used within the determination of SZ measurements themselves. Our simulations suggest that for a halo with a true $Y_{\rm 200c}\simeq 10^{-4}$, we would measure a 10\% underestimation of $Y$ by neglecting relativistic effects. We provide $\Ty-Y$ relations to allow for further detailed modelling.
\end{itemize}

Future and ongoing experiments such as CCAT-Prime \citep{Stacey2018}, NIKA2 \citep{NIKA2updated}, TolTEC \citep{TolTEC}, and The Simons Observatory \citep{TheSimonsObservatoryCollaboration2018} offer an exciting potential for measuring the ICM temperature using the rSZ effect. This will enable comprehensive analyses of the ICM structure and evolution, especially for high-redshift clusters where X-ray temperatures are difficult to obtain. These will also provide an observational test for the validity of the simulation results presented here. Where rSZ temperatures cannot be directly measured, we have provided temperature scaling relations that can be used widely to estimate the impact and potential constraining power of the relativistic SZ effects in future measurements.


{\small
\section*{Acknowledgements}
The authors would like to thank Ian McCarthy for use of the $\Bah$ and $\Mac$ simulations.
This work was supported by the ERC Consolidator Grant {\it CMBSPEC} (No.~725456).
PS gratefully acknowledges support from the YCAA Prize Postdoctoral Fellowship and helpful discussions with Antonio Ragagnin, Alex Saro and Veronica Biffi.
EL was supported by the Royal Society on grant No.~RGF/EA/180053.
DA is supported by the National Science Foundation Graduate Research Fellowship under Grant No. DGE 1746045.
JC was furthermore supported by the Royal Society as a Royal Society University Research Fellow at the University of Manchester, UK (No.~URF/R/191023).
WC is supported by the STFC AGP Grant ST/V000594/1 and the Atracci\'{o}n de Talento Contract no. 2020-T1/TIC-19882 granted by the Comunidad de Madrid in Spain. He further acknowledges the science research grants from the China Manned Space Project with NO. CMS-CSST-2021-A01 and CMS-CSST-2021-B01.
KD acknowledges support by the COMPLEX project from the European Research Council (ERC) under the European Union’s Horizon 2020 research and innovation program grant agreement ERC-2019-AdG 882679 as well as support by the Deutsche Forschungsgemeinschaft (DFG, German Research Foundation) under Germany’s Excellence Strategy - EXC-2094 - 390783311. 
GY acknowledges financial support from the MICIU/FEDER (Spain) under project grant PGC2018-094975-C21. WC and GY would like to thank Ministerio de Ciencia e Innovacion (Spain) for financial support under research grant  PID2021-122603NB-C21

This work used the DiRAC@Durham facility managed by the Institute for Computational Cosmology on behalf of the STFC DiRAC HPC Facility (www.dirac.ac.uk). The equipment was funded by BEIS capital funding via STFC capital grants ST/K00042X/1, ST/P002293/1, ST/R002371/1 and ST/S002502/1, Durham University and STFC operations grant ST/R000832/1. DiRAC is part of the National e-Infrastructure
The \textsc{Magneticum} simulations were carried out at the Leibniz Supercomputer Center (LRZ) under the project pr83li.
The Illustris$\Tng$ simulations were run with compute time granted by the Gauss Centre for Supercomputing (GCS) under Large-Scale Projects GCS-ILLU and GCS-DWAR on the GCS share of the supercomputer Hazel Hen at the High Performance Computing Center Stuttgart (HLRS). 
$\The$ project has received financial support from the European Union’s Horizon 2020 Research and Innovation programme under the Marie Sklodowskaw-Curie grant agreement number 734374, i.e. the LACEGAL project. We would like to thank The Red Espa{\~n}ola de Supercomputaci{\'o}n for granting us computing time at the MareNostrum Supercomputer of the BSC-CNS where most of the 300 cluster simulations have been performed. The MDPL2 simulation has been performed at LRZ Munich within the project pr87yi. The CosmoSim database (https://www.cosmosim.org) is a service by the Leibniz Institute for Astrophysics Potsdam (AIP). Part of the computations with \texttt{GADGET-X} have also been performed at the ‘Leibniz-Rechenzentrum’ with CPU time assigned to the Project ‘pr83li’.
}
\section*{Data Availability}
Public data releases exist for \textsc{IllustrisTNG} \citep{Nelson2019} and $\Mag$ \citep{Ragagnin2017}, and can be found at {\tt https://www.tng-project.org/} and {\tt http://magneticum.org/data.html} respectively. The thermodynamic properties presented in this work are not part of the public catalogs but can be provided on reasonable request. 
The data for $\The$, $\Bah$ and $\Mac$ are not available in a public repository, but can be provided on request.

{\small
\bibliographystyle{mnras}
\bibliography{TemperatureBias}
}

\appendix
\section{Simulation specific effects}
\label{app:Sim_effects}
In this section, we examine how simulation specific quantities modify our observed results. In particular, in Section~\ref{app:relaxation}, we examine a simplistic relaxation model across all 4 simulations. We consider different feedback models within the $\The$ simulations in Section~\ref{sec:Res_Feedback}. Section~\ref{sec:Res_Resolution} considers the effects of resolution, focusing on $\Tng$. In Section~\ref{app:core-excision} we consider varying the radius of core-excision, and the effects of core-excision in $\Mag$, and finally in Section~\ref{app:T-cut}, we study the effects of changing the particle temperature cut-off in $\Mag$ 

\vspace{-4mm}
\subsection{Relaxation}
\label{app:relaxation}
\begin{figure}
    \includegraphics[width=\linewidth]{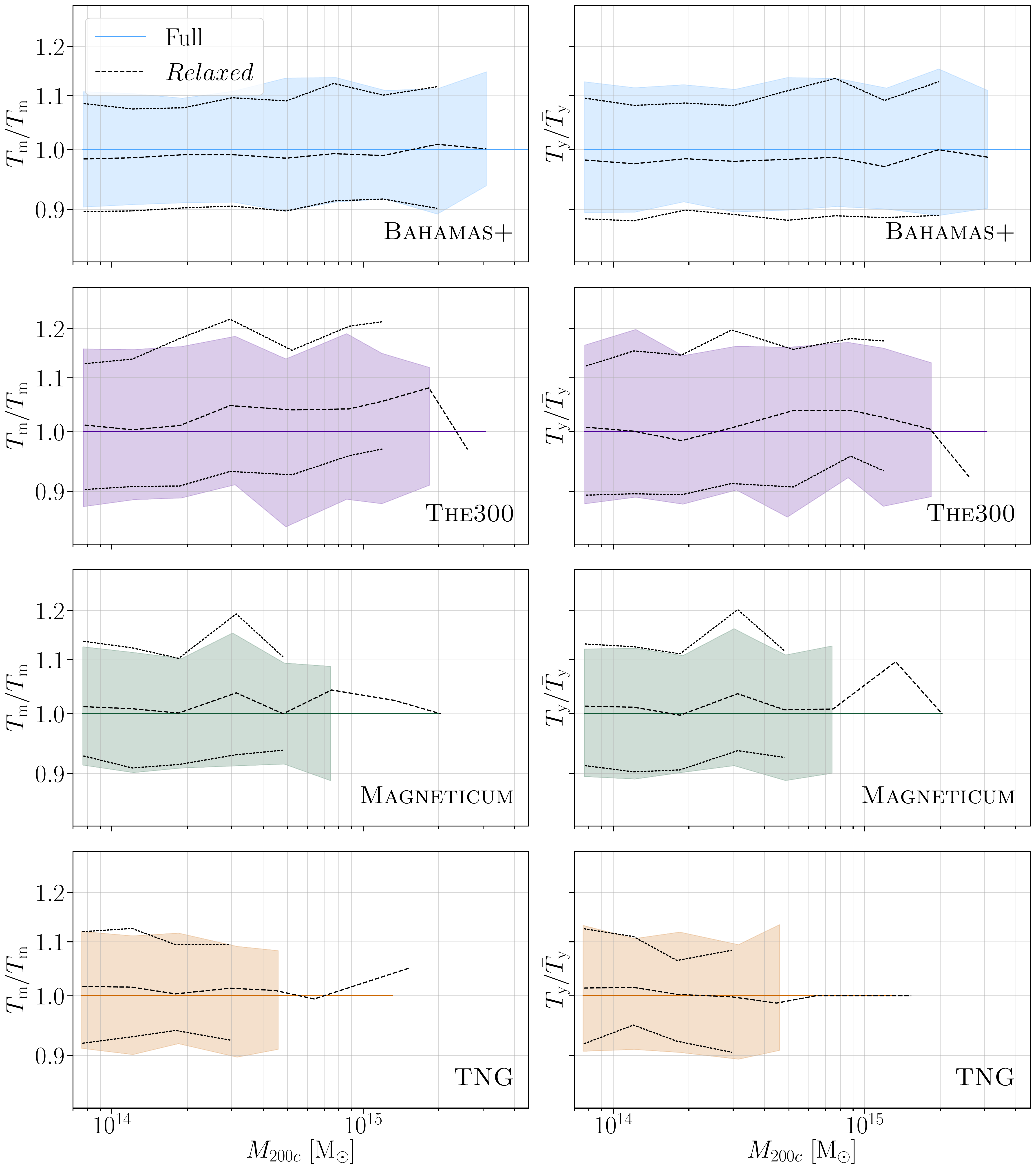}\\
    \caption{The effects of relaxation criteria on the observed temperature measures. Here the data is binned into logarithmically spaced mass bins, and the median of the whole sample is designated by $\Bar{T}$. The shaded regions show the 16 and 84 percentile regions within these bins, divided by the $\Bar{T}$. The relaxed subsample bin means (divided by $\Bar{T}$) are shown as dashed lines, while the relaxed 16 and 84 percentile regions are shown by the dotted lines. Here we can see that relaxation metrics show a limited effect on final averages of the data.}
    \label{fig:02_relaxation}
\end{figure}

It is worth briefly considering the effects on relaxation in clusters. While there are many ways to define a relaxed halo \citep[see e.g.][]{Neto2007,Duffy2008,Klypin2011,Dutton2014,Klypin2016,Barnes2017CEagle}, here we use the same criteria as defined in \citet{Henson2017}. That is, 
\begin{equation*}
	X_\mathrm{off} < 0.07;\; f_\mathrm{sub} < 0.1 \; \mathrm{and}\; \lambda < 0.07,
\end{equation*}
with $X_\mathrm{off}$ the distance offset between the point of the minimum gravitational potential in a cluster and its centre of mass, divided by $\Rvir$; $f_\mathrm{sub}$ the mass fraction within $\Rvir$ that is bound to substructures and $\lambda$ the spin parameter for all particles within $\Rtwohc$.\footnote{We note, that for $\Mag$ we have calculated the first constraints on $X_{\rm off}$ and $f_{\rm sub}$. However as this restricts the population to a similar proportion of the haloes as is obtained with all three criteria in our other samples, we consider this to be a sufficient representation of the cluster relaxation for the purposes of this paper.}
It should be noted that this is not a small sample of the most relaxed objects, but instead a simple metric to remove those that are significantly disturbed.

In Fig.~\ref{fig:02_relaxation}, we show these considerations for our higher mass clusters at $z=0$. We can immediately see that there is little evidence of a bulk offset or significant change in intrinsic temperature spread between relaxed and un-relaxed clusters. However, we do see a slight raise in the temperatures within $\The$ at high masses. This is not present at lower masses, nor meaningfully in the other simulations, however. As such, we do not consider relaxation in more detail here.

\vspace{-4mm}
\subsection{Feedback}
\label{sec:Res_Feedback}
\begin{figure}
    \includegraphics[width=\linewidth]{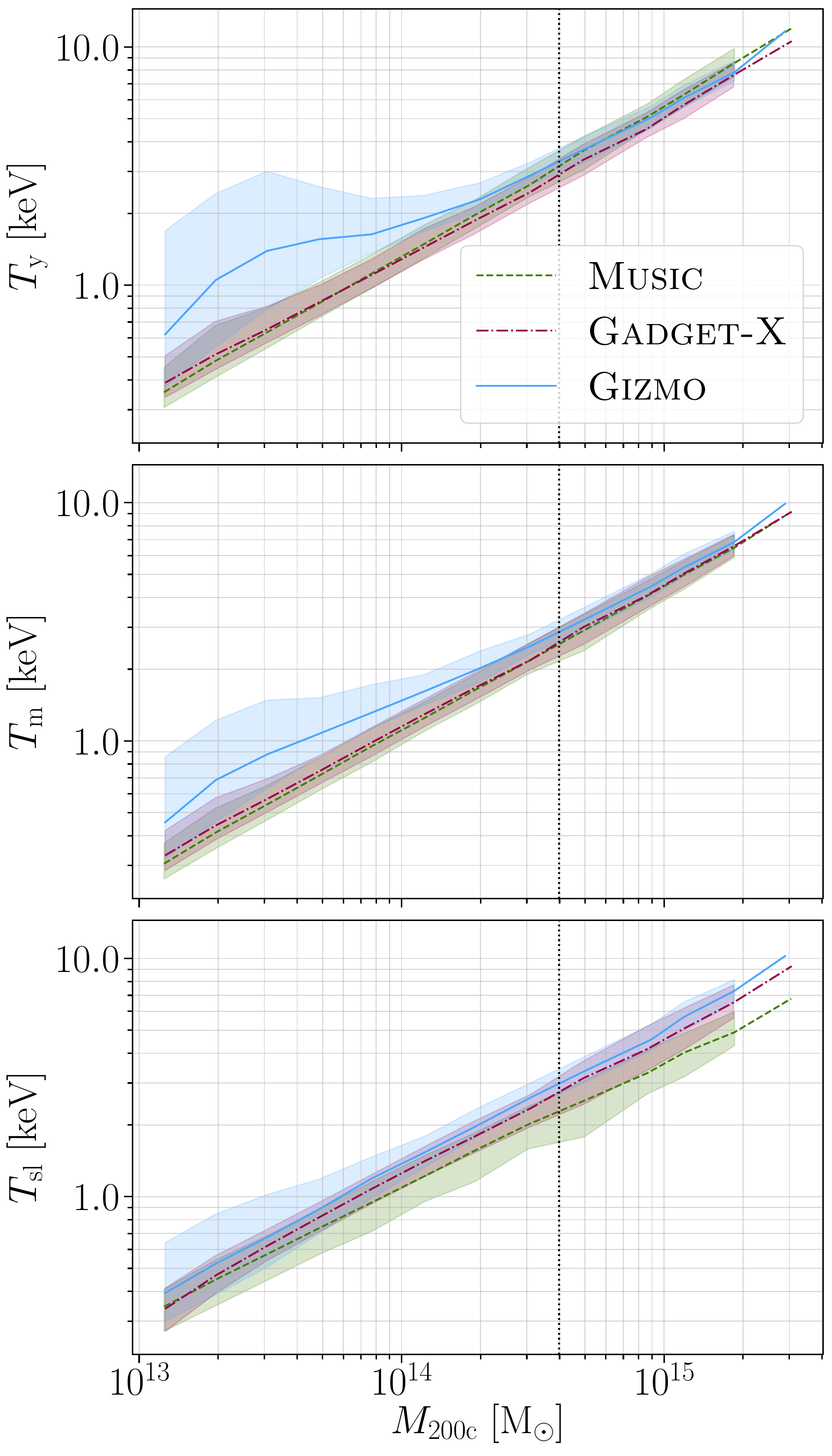}\\
    \caption{A comparison of different feedback mechanisms within The300. Here {\sc Gadget-X} is the run used in the rest of the paper, {\sc Music} is a run without any AGN feedback, and {\sc Gizmo} is a run using a mass independent feedback calibrated for the highest masses. The black vertical dotted line at $\Mtwohc=10^{14.6}$~$\msol$ indicates an approximation for the region above which The300 was predominantly calibrated. This figure is otherwise arranged as in Fig.~\ref{fig:03_TvM_z0}.}
    \label{fig:11_Feedback_The300}
\end{figure}
As discussed in Section~\ref{sec:Simulations}, all of our samples use very differently calibrated feedback models.\footnote{That is, not only are they calibrated to different properties but also, the feedback models themselves are varied between each simulation.} This makes it incredibly difficult to judge how the different feedback choices have changed our observed results. As such,  In this section, we compare the three different runs of $\The$ described above.

In Fig.~\ref{fig:11_Feedback_The300}, we show the three different $\The$ runs -- here {\sc Gadget-X} is the run used in the rest of the paper. {\sc Gizmo} is a run using a more refined feedback model that is nonetheless calibrated for high mass clusters, leading to an observed lower $\fgas$, particularly at low masses than the other runs. This is reflected in the {\sc Gizmo} temperatures lying high at low masses. {\sc Music} on the other hand has no AGN feedback, and accordingly a shallower slope in $\fgas$ than {\sc Gadget-X}. It is worth noting that due to the calibration regimes, {\sc Music} thus has slightly higher values for $\fgas$ at low masses and lower values, at high masses.

Examining $\Ty$ we can immediately see a reflection of the feedback. At high masses where $\The$ is predominantly calibrated, {\sc Gizmo} and {\sc Gadget-X} agree, while {\sc Gizmo} lies high as previously noted for lower masses. {\sc Music} lies a little low at the lowest masses compared to {\sc Gadget-X} and high at the highest -- the divergence between the three runs appears to happen almost inversely proportionally to the variation in $\fgas$ between the three runs. This matches our observations for $\fgas$ from the previous section.

Similar, but far reduced effects can be seen by examining $\Tm$. We find that at high masses, all three feedback runs, agree extraordinarily well. This implies that large variations in feedback -- e.g., the large feedback at low masses in {\sc Gizmo} -- 
are required to disrupt the stability of $\Tm$.

Examining $\Tsl$ we do however see very different behaviour. Here the three runs diverge significantly. {\sc Gizmo} lies almost at a constant increase above {\sc Gadget-X}, {\sc Music}, on the other hand, has a significantly shallower slope than the other two runs, with a large decrease in $\Tsl$ at high masses. This indicates that $\Tsl$ is somewhat more affected by variations in feedback models.

\vspace{-4mm}
\subsection{Resolution}
\label{sec:Res_Resolution}
\begin{figure}
    \includegraphics[width=\linewidth]{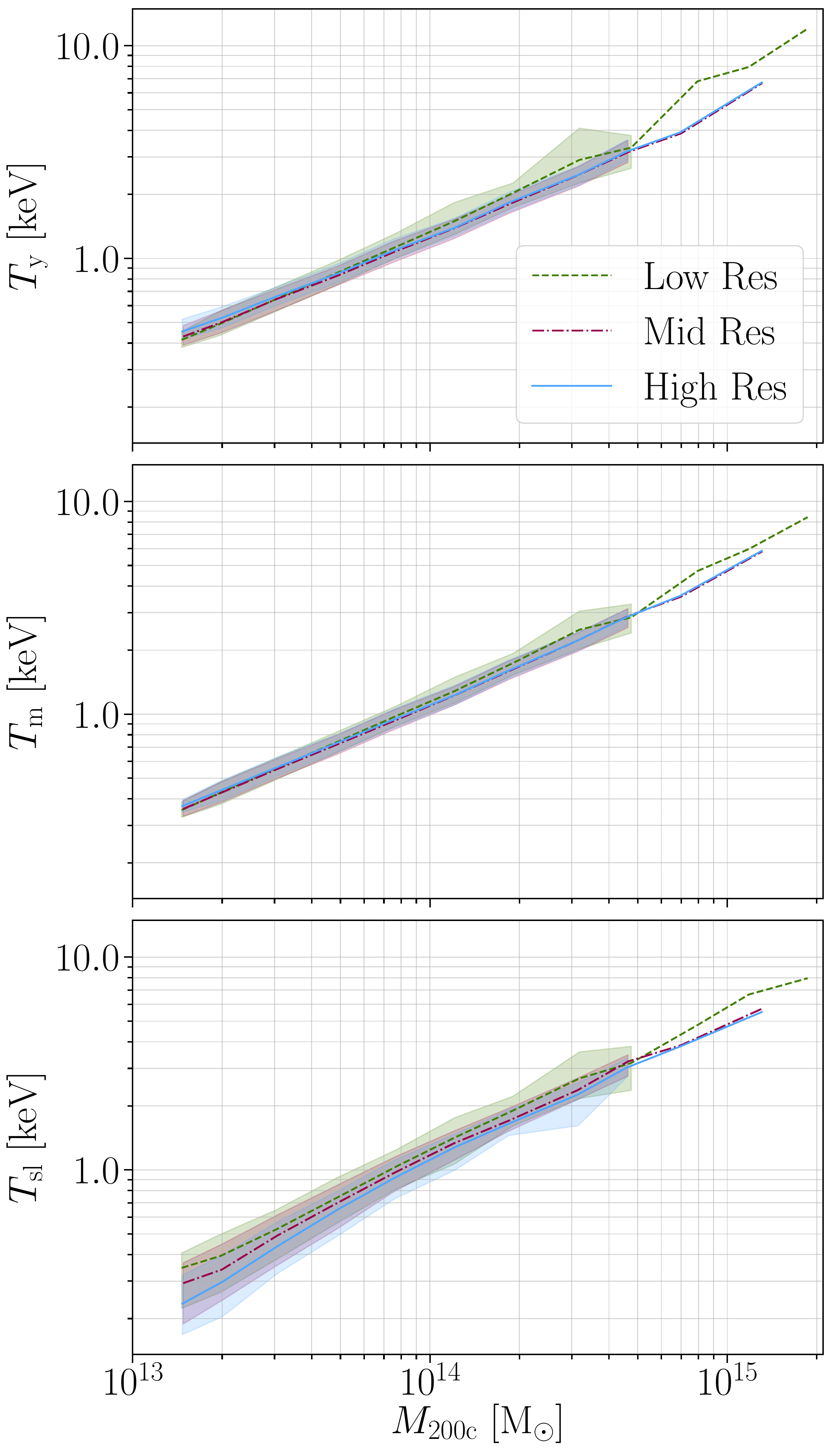}\\
    \caption{The effects of resolution in the TNG simulation. High Res here refers to the run used within the rest of this paper. Mid Res and Low Res are runs with x8 and x64 worse resolution respectively. This figure is arranged as in Fig.~\ref{fig:03_TvM_z0}.}
    \label{fig:12_Resolution_TNG}
\end{figure}
Resolution studies are in general complicated as many simulations do not have different runs at different resolutions. However, here we can consider 3 different runs of $\Tng$ at different resolutions, alongside briefly discussing the two different boxes from $\Mag$.

The effects of the 3 different resolutions in $\Tng$ can be seen in Fig.~\ref{fig:12_Resolution_TNG}. Here the highest resolution is the one we have used throughout the rest of this work, while the other two have 8 and 64 times worse resolutions respectively. In $\Tng$, we can see that resolution has little effect on the SZ temperatures ($\Tm$ and $\Ty$), with differences only occurring in the lowest resolution run at high masses, where the temperatures tend to fall a little higher.

We see more systematic differences in $\Tsl$ however, where with increasing resolution, the temperatures all seem to fall, especially in the lower mass haloes. This is likely due to the cold dense clumps in the haloes being better resolved in the higher resolution runs and as such, leading to lower averaged values for $\Tsl$.

We also compare the $\Mag$ {\tt Box2 hr} run we have been using throughout the rest of this work, with the {\tt Box1a mr}, a run with around $\simeq 20$ times worse resolution. Here we do see a decrease for both $\Tm$ and $\Ty$ in the lower resolution study, of a similar scale to the intrinsic scatter. However, it is also worth noting that these two runs do have different subgrid physics, as they are independently calibrated, which may contribute to this difference. Due to the scatter inherent in the $\Mag$ values for $\Tsl$ it is hard to gain a conclusive indication of how these vary.

We can then consider the differences between simulations in the formalism discussed in \citet{Schaye2015}, as a `strong' convergence test using $\Tng$ and a `weak' convergence test using $\Mag$. As such, the small variation that is observed in $\Mag$ is likely caused by the recalibration, more than the inherent resolution variation itself, given the lack of any resolution dependence in $\Tng$. However, more detailed work may be necessary to determine how in general resolution effects may change the temperatures gained from simulations.

\vspace{-4mm}
\subsection{Core-Excision}
\label{app:core-excision}
\begin{figure}
    \includegraphics[width=0.9\linewidth]{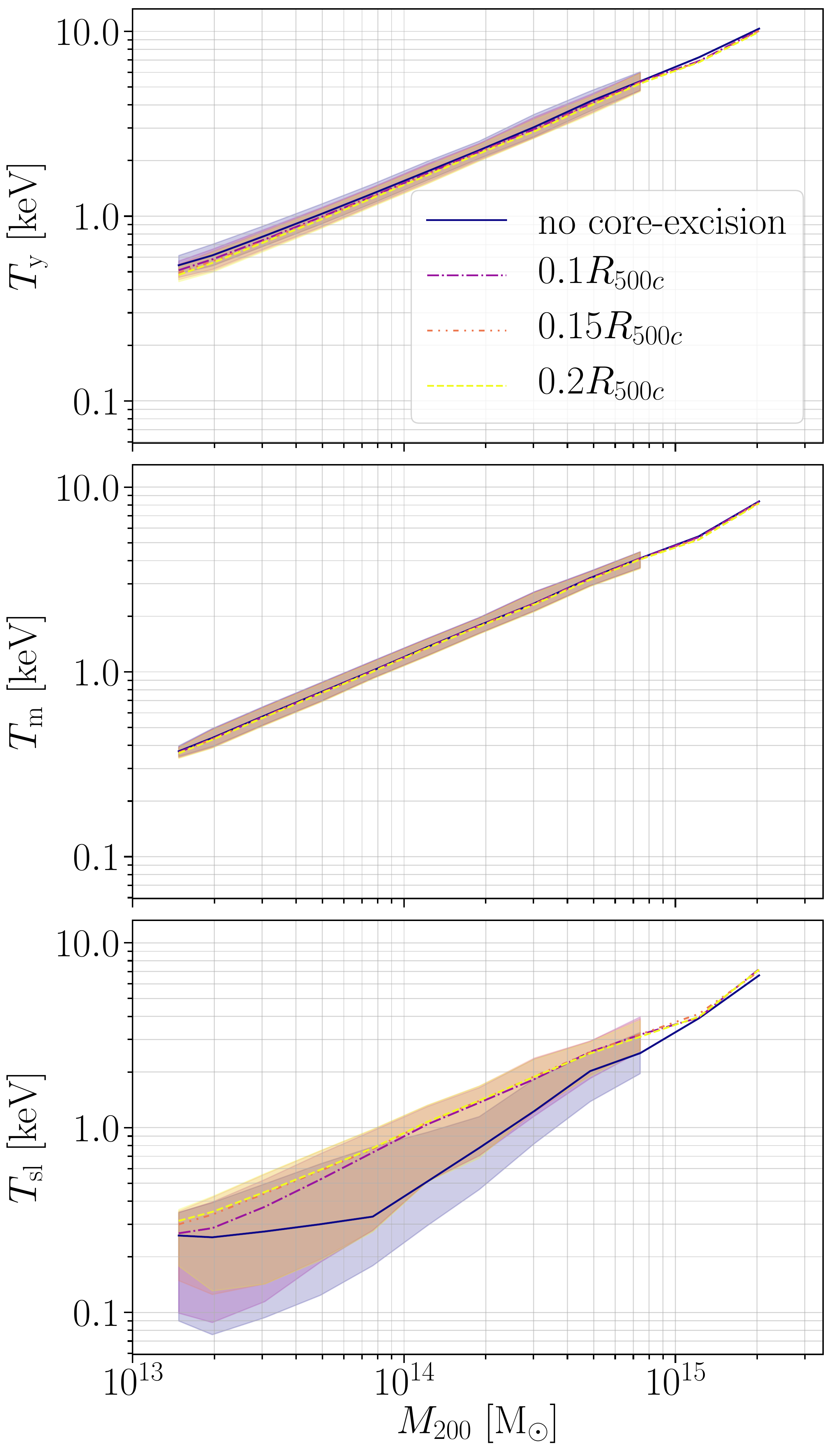}\\
    \caption{The effects of core excision on the observed temperature measures within the $\Mag$ sample at $z=0$. The line labelled `no CE' is the volume averages with no core-excision, while the other lines are marked with the radius of the core-excision (i.e., any cells within this radius are ignored in the averaging procedure). We see little change to the value of $\Tm$ but a slight decrease within $\Ty$. At higher redshifts, this effect is magnified within the $\Mag$ sample. However, we note that by core-excision of around $0.15\Rfivehc$ these effects have become stable.}
    \label{fig:01_core_excision}
\end{figure}

The $\Mag$ simulation, in contrast to our other simulations, shows a large variation in the $y$-weighted temperature measure averaged over the whole halo volume, especially at higher redshifts. This appears to be due to a difference in the gas behaviour of the core in the $\Mag$ simulations compared to the other simulations ($\BaM$, $\The$, and $\Tng$), and can be mitigated with core-excision.

Within the $\Mag$ simulation we find a far larger proportion of hot ($>10^8$~K) gas cells in the clusters than in our other simulations, particularly located in the core of the clusters. We note that in general, none of the simulations model the cluster cores with high fidelity, however only within the $\Mag$ simulations do we see these core effects significantly affecting our averaged SZ temperatures. In particular, in the $\Mag$ sample, at $z=1.5$ within $\Rfivehc$, most clusters have a small fraction ($\lesssim 0.005$\%) of the gas cells having these very high temperatures, with a maximal fraction of 2\% in the most extreme cluster. The TNG sample, in contrast, has a tiny fraction -- for most clusters $\lesssim 0.0005$\% of the cells (i.e., a factor of 10 fewer), with a maximal fraction of 0.5\% in the most extreme cluster.

These very high-temperature gas cells will upweight the averaged $\Ty$ more than $\Tm$ leading to the observed changes. And, due to both the instabilities within the $\Ty$ data predominantly being generated at higher redshifts and the lack of these similar effects within the other simulations, we assume that this effect is unphysical.

In Fig.~\ref{fig:01_core_excision}, we can see the effects of different radii of core-excision within the $\Mag$ data. In particular, we can motivate the use of the same core-excision used for our $\Tsl$ values, i.e., $0.15\Rfivehc$, as this indicates the turning point, after which further core-excision shows little change. This also, at higher redshifts, results in removing the unphysical variation present in the $\Ty$ data.

As such, we use the core-excised values for all three temperature measures ($\Ty$, $\Tm$, and $\Tsl$) within the $\Mag$ sample. For our other three samples, we use this core-excision {\it only} for $\Tsl$ and use the whole cluster averages for $\Ty$ and $\Tm$.

\vspace{-4mm}
\subsection{Particle temperature cut}
\label{app:T-cut}
\begin{figure}
    \includegraphics[width=0.9\linewidth]{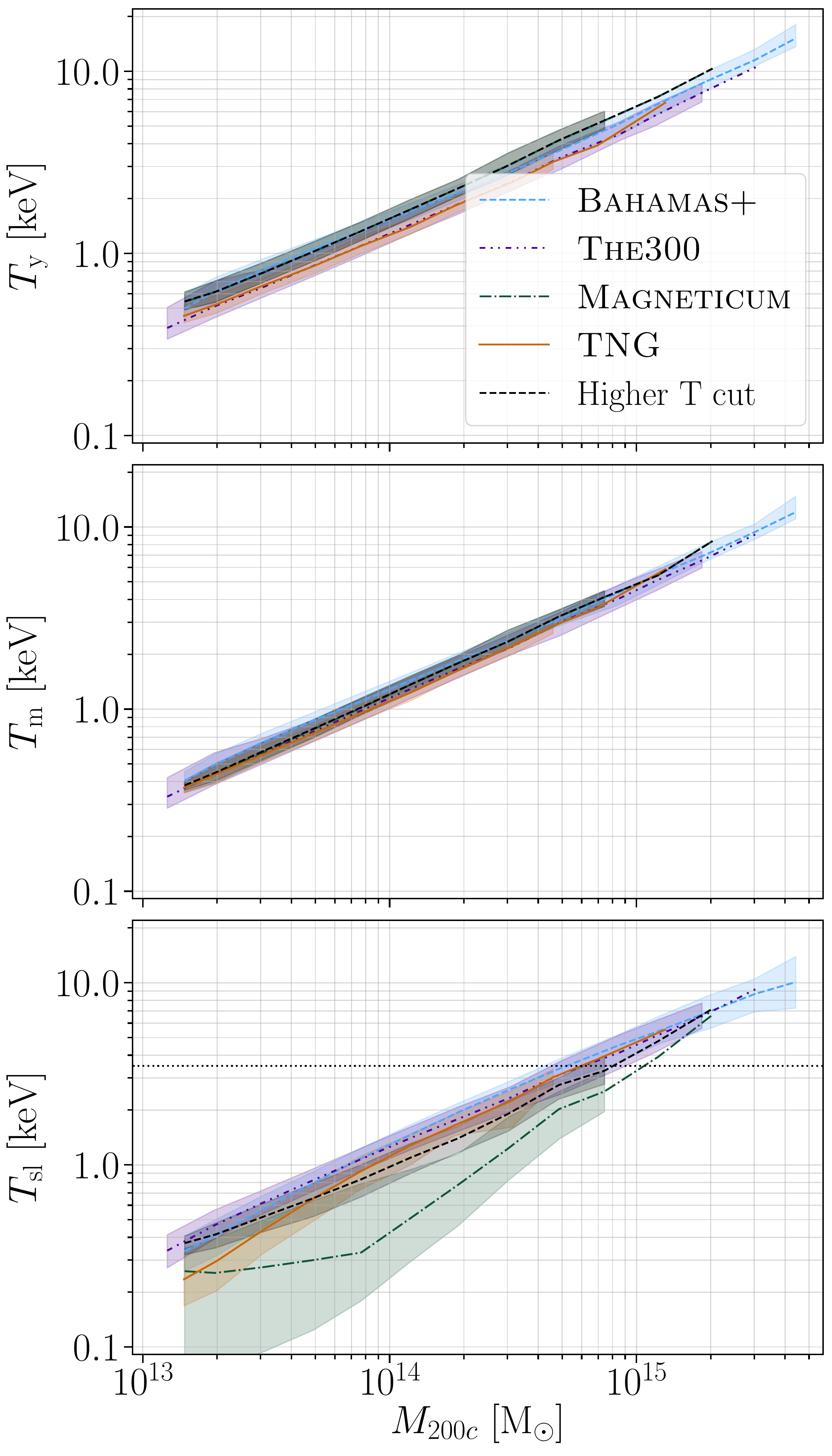}\\
    \caption{The effects of the temperature cut off on the $\Mag$ sample at $z=0$. The line labelled `Higher T cut' is the $\Mag$ volume averages using a temperature cut of $T\geq10^6$~K. Otherwise, the figure is as described in Fig.~\ref{fig:03_TvM_z0}.}
    \label{fig:03a_Tcut}
\end{figure}

In Section~\ref{sec:Simulations}, we indicate that as is a common practice we use a temperature cut to exclude low-temperature particles that wouldn't emit X-rays, and would otherwise bias $\Tsl$. In particular, we use a temperature cut of $T\geq10^{5.2}$~K.
However, in Section~\ref{sec:TS}, we observe large scatter in the $\Tsl$ values found in $\Mag$. These seem to be driven in large part by warm particles, not excluded in this regime.

That is, in Figure~\ref{fig:03a_Tcut}, we have plotted the four samples we discuss in the body of this paper, alongside the $\Mag$ haloes when a stricter temperature cut $T\geq10^{6}$~K is implemented. We can immediately see that the scatter in $\Tsl$ is greatly reduced, and the medians lie far closer to those obtained from the other simulations. These warm particles account for around 1\% of the particles in most haloes, increasing at lower masses. In some of the lowest mass haloes these warm particles account for around $\simeq10$\% of the particles. As such, we do not use this temperature cut in $\Mag$ globally, but use the consistent temperature cut of $T\geq10^{5.2}$~K across all simulations.

We can also see that, at least in $\Mag$, changing the temperature cut seems to have very minimal effects on $\Ty$ and $\Tm$. This is to be expected, as these are far more weighted by the higher temperature gas components in haloes. In principle, the SZ effect will be caused to some extent by {\it all} the gas in a halo, not merely the hottest gasses, however this consistency across temperature cuts, shows that the contribution of the warm gas to the global SZ signal is minimal. This indicates that $\Tm$ and $\Ty$ are robust against the choice of temperature cut.

\vspace{-4mm}
\section{Fits}
\label{app:tables}
\begin{table}
\caption{The two parameter fits [Eq.~\eqref{eqn:fit_format}] and a measure of the intrinsic scatter [Eq.~\eqref{eqn:sigma_def}] in the fits at $z=0$ for the median and 16 and 84 percentiles of the cross simulation averaged sample within $\Rtwohc$.} \centering
\begin{tabular}{lccc}
 median & $A$ & $B$ & $\slog$ \\ 
 \hline \smallskip
 $\Ty$  & $1.465^{+0.002}_{-0.002}$ & $0.586^{+0.003}_{-0.002}$ & $0.1025^{+0.0003}_{-0.0002}$\\ \smallskip
 $\Tm$  & $1.210^{+0.001}_{-0.001}$ & $0.591^{+0.003}_{-0.003}$ & $0.0805^{+0.0002}_{-0.0001}$\\ \smallskip
 $\Tsl$ & $1.135^{+0.003}_{-0.003}$ & $0.601^{+0.006}_{-0.011}$ & $0.2067^{+0.0016}_{-0.0008}$ \\
 \hline
 16\% &&& \\
 \hline \smallskip
 $\Ty$  & $1.259^{+0.002}_{-0.003}$ & $0.590^{+0.002}_{-0.002}$ & --\\ \smallskip
 $\Tm$  & $1.063^{+0.002}_{-0.002}$ & $0.597^{+0.002}_{-0.002}$ & --\\ \smallskip
 $\Tsl$ & $0.872^{+0.003}_{-0.003}$ & $0.636^{+0.003}_{-0.003}$ & --\\
 \hline
 84\% &&& \\
 \hline \smallskip
 $\Ty$  & $1.694^{+0.003}_{-0.003}$ & $0.573^{+0.002}_{-0.002}$ & --\\ \smallskip
 $\Tm$  & $1.381^{+0.002}_{-0.002}$ & $0.585^{+0.003}_{-0.002}$ & --\\ \smallskip
 $\Tsl$ & $1.364^{+0.003}_{-0.003}$ & $0.611^{+0.002}_{-0.002}$ & --\\
 \hline
\end{tabular}
\label{apptab:fits_median_z0_all}
\end{table}
\begin{table}
\caption{The three parameter fits [Eq.~\eqref{eqn:fit_format}] and a measure of the intrinsic scatter [Eq.~\eqref{eqn:sigma_def}] in the fits at $z=0$ for the median and 16 and 84 percentiles within the cross simulation averaged sample within $\Rtwohc$.} \centering
\begin{tabular}{lcccc}
 median & $A$ & $B$ & $C$ & $\slog$ \\ 
 \hline \smallskip
 $\Ty$  & $1.426^{+0.006}_{-0.007}$ & $0.566^{+0.001}_{-0.001}$ & $0.024^{+0.005}_{-0.004}$ & $0.1011^{+0.0001}_{-0.0001}$\\ \smallskip
 $\Tm$  & $1.207^{+0.005}_{-0.005}$ & $0.589^{+0.001}_{-0.001}$ & $0.003^{+0.004}_{-0.005}$ & $0.0804^{+0.0001}_{-0.0001}$\\ \smallskip
 $\Tsl$ & $1.196^{+0.020}_{-0.009}$ & $0.641^{+0.003}_{-0.003}$ & $-0.048^{+0.007}_{-0.018}$ & $0.2028^{+0.0002}_{-0.0001}$\\
 \hline
 16\% &&& \\
 \hline \smallskip
 $\Ty$  & $1.248^{+0.005}_{-0.005}$ & $0.585^{+0.001}_{-0.001}$ & $0.008^{+0.004}_{-0.004}$ & --\\ \smallskip
 $\Tm$  & $1.068^{+0.004}_{-0.003}$ & $0.600^{+0.001}_{-0.001}$ & $-0.005^{+0.003}_{-0.003}$ & --\\ \smallskip
 $\Tsl$ & $0.920^{+0.006}_{-0.005}$ & $0.667^{+0.003}_{-0.003}$ & $-0.051^{+0.004}_{-0.005}$ & --\\
 \hline
 84\% &&& \\
 \hline \smallskip
 $\Ty$  & $1.641^{+0.005}_{-0.004}$ & $0.555^{+0.001}_{-0.001}$ & $0.030^{+0.002}_{-0.003}$ & --\\ \smallskip
 $\Tm$  & $1.370^{+0.004}_{-0.005}$ & $0.580^{+0.001}_{-0.001}$ & $0.007^{+0.004}_{-0.002}$ & --\\ \smallskip
 $\Tsl$ & $1.398^{+0.005}_{-0.005}$ & $0.625^{+0.002}_{-0.002}$ & $-0.023^{+0.003}_{-0.003}$ & --\\
 \hline
\end{tabular}
\label{apptab:fits_3_median_z0_all}
\end{table}
\begin{table}
\caption{The two and three parameter fits [Eq.~\eqref{eqn:fit_format}] and a measure of the intrinsic scatter [Eq.~\eqref{eqn:sigma_def}] in the fits at $z=0$ for the median of each sample within $\Rtwohc$.} \centering
\begin{tabular}{lcccc}
 \multicolumn{5}{c}{$\BaM$} \\
 & $A$ & $B$ & C & $\slog$ \\ 
 \hline \smallskip
 $\Ty$  & $1.529^{+0.003}_{-0.003}$ & $0.582^{+0.003}_{-0.003}$ & 0 & $0.0829^{+0.0005}_{-0.0004}$\\ \smallskip
 $\Ty$  & $1.494^{+0.007}_{-0.008}$ & $0.564^{+0.002}_{-0.002}$ & $0.022^{+0.005}_{-0.004}$ & $0.0808^{+0.0001}_{-0.0001}$\\ \smallskip
 $\Tm$  & $1.264^{+0.002}_{-0.002}$ & $0.581^{+0.003}_{-0.003}$ & 0 & $0.0445^{+0.0002}_{-0.0002}$\\ \smallskip
 $\Tm$  & $1.261^{+0.006}_{-0.006}$ & $0.580^{+0.002}_{-0.002}$ & $0.002^{+0.004}_{-0.005}$ & $0.0444^{+0.0001}_{-0.0001}$\\ \smallskip
 $\Tsl$ & $1.186^{+0.004}_{-0.004}$ & $0.596^{+0.007}_{-0.011}$ & 0 & $0.0945^{+0.0024}_{-0.0013}$\\ \smallskip
 $\Tsl$ & $1.277^{+0.022}_{-0.010}$ & $0.654^{+0.004}_{-0.003}$ & $-0.069^{+0.008}_{-0.019}$ & $0.0849^{+0.0002}_{-0.0001}$\\
 \hline
 \multicolumn{5}{c}{$\The$} \\
 & $A$ & $B$ & $C$ & $\slog$ \\ 
 \hline \smallskip
 $\Ty$  & $1.295^{+0.004}_{-0.004}$ & $0.596^{+0.005}_{-0.005}$ & 0 & $0.1779^{+0.0008}_{-0.0008}$\\ \smallskip
 $\Ty$  & $1.263^{+0.010}_{-0.011}$ & $0.582^{+0.002}_{-0.002}$ & $0.023^{+0.009}_{-0.009}$ & $0.1754^{+0.0003}_{-0.0003}$\\ \smallskip
 $\Tm$  & $1.137^{+0.004}_{-0.004}$ & $0.599^{+0.005}_{-0.004}$ & 0 & $0.1567^{+0.0007}_{-0.0006}$\\ \smallskip
 $\Tm$  & $1.125^{+0.009}_{-0.009}$ & $0.593^{+0.002}_{-0.002}$ & $0.010^{+0.009}_{-0.008}$ & $0.1558^{+0.0003}_{-0.0002}$\\ \smallskip
 $\Tsl$ & $1.205^{+0.004}_{-0.006}$ & $0.585^{+0.005}_{-0.012}$ & 0 & $0.1520^{+0.0005}_{-0.0001}$\\ \smallskip
 $\Tsl$ & $1.224^{+0.020}_{-0.011}$ & $0.593^{+0.002}_{-0.002}$ & $-0.013^{+0.007}_{-0.022}$ & $0.1513^{+0.0003}_{-0.0005}$\\
 \hline
 \multicolumn{5}{c}{$\Mag$} \\
 & $A$ & $B$ & $C$ & $\slog$ \\ 
 \hline \smallskip
 $\Ty$  & $1.511^{+0.007}_{-0.006}$ & $0.612^{+0.006}_{-0.003}$ & 0 & $0.0452^{+0.0005}_{-0.0003}$\\ \smallskip
 $\Ty$  & $1.473^{+0.006}_{-0.005}$ & $0.601^{+0.003}_{-0.003}$ & $0.026^{+0.006}_{-0.004}$ & $0.0438^{+0.0000}_{-0.0000}$\\ \smallskip
 $\Tm$  & $1.182^{+0.004}_{-0.004}$ & $0.622^{+0.004}_{-0.004}$ & 0 & $0.0327^{+0.0002}_{-0.0001}$\\ \smallskip 
 $\Tm$  & $1.167^{+0.004}_{-0.004}$ & $0.616^{+0.003}_{-0.003}$ & $0.013^{+0.004}_{-0.004}$ & $0.0326^{+0.0001}_{-0.0001}$\\ \smallskip
 $\Tsl$ & $0.930^{+0.007}_{-0.006}$ & $0.629^{+0.008}_{-0.006}$ & 0 & $0.3020^{+0.0008}_{-0.0008}$\\ \smallskip
 $\Tsl$ & $0.904^{+0.008}_{-0.008}$ & $0.616^{+0.006}_{-0.006}$ & $0.030^{+0.007}_{-0.006}$ & $0.3035^{+0.0007}_{-0.0007}$\\
 \hline
 \multicolumn{5}{c}{$\Tng$} \\
 & $A$ & $B$ & $C$ & $\slog$ \\ 
 \hline \smallskip
 $\Ty$  & $1.302^{+0.008}_{-0.008}$ & $0.583^{+0.007}_{-0.008}$ & 0 & $0.0388^{+0.0010}_{-0.0010}$\\ \smallskip
 $\Ty$  & $1.242^{+0.011}_{-0.010}$ & $0.569^{+0.005}_{-0.005}$ & $0.057^{+0.012}_{-0.014}$ & $0.0356^{+0.0002}_{-0.0002}$\\ \smallskip
 $\Tm$  & $1.129^{+0.006}_{-0.007}$ & $0.604^{+0.005}_{-0.008}$ & 0 & $0.0322^{+0.0007}_{-0.0008}$\\ \smallskip
 $\Tm$  & $1.089^{+0.008}_{-0.007}$ & $0.594^{+0.004}_{-0.005}$ & $0.042^{+0.009}_{-0.012}$ & $0.0304^{+0.0002}_{-0.0002}$\\ \smallskip
 $\Tsl$ & $0.998^{+0.008}_{-0.009}$ & $0.704^{+0.009}_{-0.009}$ & 0 & $0.2030^{+0.0009}_{-0.0008}$\\ \smallskip
 $\Tsl$ & $1.070^{+0.012}_{-0.013}$ & $0.722^{+0.008}_{-0.008}$ & $-0.084^{+0.019}_{-0.013}$ & $0.2010^{+0.0006}_{-0.0005}$\\
 \hline
\end{tabular}
\label{apptab:fits_median_z0_samples}
\end{table}

\begin{table}
\caption{The two parameter fits [Eq.~\eqref{eqn:fit_format}] and a measure of the intrinsic scatter [Eq.~\eqref{eqn:sigma_def}] in the fits for the median across the higher redshifts of the cross simulation averaged sample within $\Rtwohc$.} \centering
\begin{tabular}{lccc}
 $z=0.25$ & $A$ & $B$ & $\slog$ \\ 
 \hline \smallskip
 $\Ty$  & $1.448^{+0.002}_{-0.002}$ & $0.600^{+0.002}_{-0.004}$ & $0.0896^{+0.0003}_{-0.0005}$\\ \smallskip
 $\Tm$  & $1.173^{+0.002}_{-0.002}$ & $0.601^{+0.004}_{-0.005}$ & $0.0669^{+0.0003}_{-0.0003}$\\ \smallskip
 $\Tsl$ & $1.059^{+0.004}_{-0.004}$ & $0.608^{+0.013}_{-0.017}$ & $0.2151^{+0.0025}_{-0.0017}$\\
 \hline
 $z=0.50$ &&& \\
 \hline \smallskip
 $\Ty$  & $1.423^{+0.003}_{-0.003}$ & $0.591^{+0.003}_{-0.002}$ & $0.0804^{+0.0002}_{-0.0002}$\\ \smallskip
 $\Tm$  & $1.136^{+0.002}_{-0.002}$ & $0.599^{+0.003}_{-0.002}$ & $0.0589^{+0.0001}_{-0.0001}$\\ \smallskip
 $\Tsl$ & $1.002^{+0.003}_{-0.003}$ & $0.633^{+0.004}_{-0.003}$ & $0.2246^{+0.0003}_{-0.0004}$\\
 \hline
 $z=1.00$ \\
 \hline \smallskip
 $\Ty$  & $1.376^{+0.004}_{-0.004}$ & $0.594^{+0.003}_{-0.003}$ & $0.0733^{+0.0001}_{-0.0001}$\\ \smallskip
 $\Tm$  & $1.073^{+0.003}_{-0.002}$ & $0.607^{+0.002}_{-0.002}$ & $0.0513^{+0.0001}_{-0.0001}$\\ \smallskip
 $\Tsl$ & $0.916^{+0.003}_{-0.003}$ & $0.661^{+0.003}_{-0.004}$ & $0.2394^{+0.0003}_{-0.0003}$\\
 \hline
 $z=1.50$ \\
 \hline \smallskip
 $\Ty$  & $1.343^{+0.006}_{-0.006}$ & $0.594^{+0.004}_{-0.004}$ & $0.0882^{+0.0001}_{-0.0001}$\\ \smallskip
 $\Tm$  & $1.038^{+0.004}_{-0.005}$ & $0.619^{+0.003}_{-0.004}$ & $0.0515^{+0.0001}_{-0.0001}$\\ \smallskip
 $\Tsl$ & $0.864^{+0.005}_{-0.005}$ & $0.682^{+0.005}_{-0.005}$ & $0.2999^{+0.0003}_{-0.0003}$\\
 \hline
\end{tabular}
\label{apptab:fits_median_Redshift_all}
\end{table}

\begin{table}
\caption{The shifts and errors in the radius, mass, $\Tm$ and $\Ty$ for each sample against those values within $\Rtwohc$. We have also repeated the cross-simulation averages found in Table~\ref{tab:R_err_all}. These values are calculated on a cluster by cluster basis, and then the averages are found within these. The central value here is the median with the errors given by the 16 and 84 percentiles. NB. We do not here have values for $\Mvir$ and $\Mtwohm$ within the TNG sample.} \centering
\begin{tabular}{lcccc}
 \multicolumn{5}{c}{Cross simulation averaged sample} \\ \smallskip
 & $R/\Rtwohc$ & $M/\Mtwohc$ & $\Tm/T_{\rm m,200c}$ & $\Ty/T_{\rm m,200c}$ \\ \smallskip
 $\Rfivehc$  & $0.66^{+0.01}_{-0.02}$ & $0.71^{+0.05}_{-0.07}$ & $1.20^{+0.04}_{-0.07}$ & $1.40^{+0.16}_{-0.12}$\\  \smallskip 
 $\Rtwohc$  & $1.00$ & $1.00$ & $1.00$ & $1.22^{+0.11}_{-0.07}$\\  \smallskip 
 $\Rfivehm$  & $1.11^{+0.01}_{-0.01}$ & $1.08^{+0.03}_{-0.02}$ & $0.95^{+0.01}_{-0.01}$ & $1.18^{+0.10}_{-0.06}$\\  \smallskip 
 $\Rvir$ & $1.33^{+0.03}_{-0.02}$ & $1.22^{+0.08}_{-0.05}$ & $0.87^{+0.04}_{-0.02}$ & $1.11^{+0.09}_{-0.06}$\\  \smallskip 
 $\Rtwohm$ & $1.64^{+0.06}_{-0.04}$ & $1.39^{+0.15}_{-0.10}$ & $0.79^{+0.07}_{-0.04}$ & $1.05^{+0.09}_{-0.06}$\\
 \hline
 \multicolumn{5}{c}{$\BaM$} \\ \smallskip
 & $R/\Rtwohc$ & $M/\Mtwohc$ & $\Tm/T_{\rm m,200c}$ & $\Ty/T_{\rm m,200c}$ \\ \smallskip
 $\Rfivehc$  & $0.66^{+0.01}_{-0.02}$ & $0.73^{+0.05}_{-0.07}$ & $1.19^{+0.05}_{-0.06}$ & $1.40^{+0.14}_{-0.09}$\\  \smallskip 
 $\Rtwohc$  & $1.00$ & $1.00$ & $1.00$ & $1.21^{+0.11}_{-0.05}$\\  \smallskip 
 $\Rfivehm$  & $1.11^{+0.01}_{-0.01}$ & $1.07^{+0.03}_{-0.02}$ & $0.95^{+0.01}_{-0.01}$ & $1.17^{+0.10}_{-0.05}$\\  \smallskip 
 $\Rvir$ & $1.33^{+0.03}_{-0.02}$ & $1.21^{+0.08}_{-0.05}$ & $0.87^{+0.03}_{-0.02}$ & $1.10^{+0.09}_{-0.04}$\\  \smallskip 
 $\Rtwohm$ & $1.63^{+0.06}_{-0.04}$ & $1.38^{+0.14}_{-0.09}$ & $0.78^{+0.05}_{-0.03}$ & $1.03^{+0.09}_{-0.05}$\\ 
 \hline
 \multicolumn{5}{c}{$\The$} \\ \smallskip
 & $R/\Rtwohc$ & $M/\Mtwohc$ & $\Tm/T_{\rm m,200c}$ & $\Ty/T_{\rm m,200c}$ \\ \smallskip$\Rfivehc$  & $0.66^{+0.01}_{-0.03}$ & $0.71^{+0.05}_{-0.08}$ & $1.16^{+0.06}_{-0.09}$ & $1.29^{+0.11}_{-0.11}$\\  \smallskip 
 $\Rtwohc$  & $1.00$ & $1.00$ & $1.00$ & $1.15^{+0.08}_{-0.04}$\\  \smallskip 
 $\Rfivehm$  & $1.12^{+0.01}_{-0.01}$ & $1.08^{+0.03}_{-0.03}$ & $0.96^{+0.02}_{-0.02}$ & $1.11^{+0.08}_{-0.04}$\\  \smallskip 
 $\Rvir$ & $1.33^{+0.03}_{-0.02}$ & $1.21^{+0.09}_{-0.06}$ & $0.90^{+0.06}_{-0.04}$ & $1.07^{+0.11}_{-0.04}$\\  \smallskip 
 $\Rtwohm$ & $1.65^{+0.06}_{-0.04}$ & $1.38^{+0.16}_{-0.11}$ & $0.84^{+0.11}_{-0.06}$ & $1.02^{+0.17}_{-0.05}$\\ 
 \hline
 \multicolumn{5}{c}{$\Mag$} \\ \smallskip
 & $R/\Rtwohc$ & $M/\Mtwohc$ & $\Tm/T_{\rm m,200c}$ & $\Ty/T_{\rm m,200c}$ \\ \smallskip$\Rfivehc$  & $0.65^{+0.01}_{-0.02}$ & $0.70^{+0.05}_{-0.07}$ & $1.22^{+0.03}_{-0.04}$ & $1.51^{+0.13}_{-0.10}$\\  \smallskip 
 $\Rtwohc$  & $1.00$ & $1.00$ & $1.00$ & $1.28^{+0.09}_{-0.06}$\\  \smallskip 
 $\Rfivehm$  & $1.12^{+0.01}_{-0.01}$ & $1.09^{+0.03}_{-0.02}$ & $0.95^{+0.01}_{-0.01}$ & $1.23^{+0.08}_{-0.05}$\\  \smallskip 
 $\Rvir$ & $1.34^{+0.03}_{-0.02}$ & $1.24^{+0.08}_{-0.05}$ & $0.87^{+0.02}_{-0.02}$ & $1.16^{+0.07}_{-0.04}$\\  \smallskip 
 $\Rtwohm$ & $1.66^{+0.06}_{-0.04}$ & $1.43^{+0.15}_{-0.10}$ & $0.78^{+0.04}_{-0.03}$ & $1.09^{+0.07}_{-0.04}$\\
 \hline
 \multicolumn{5}{c}{$\Tng$} \\ \smallskip
 & $R/\Rtwohc$ & $M/\Mtwohc$ & $\Tm/T_{\rm m,200c}$ & $\Ty/T_{\rm m,200c}$ \\ \smallskip$\Rfivehc$  & $0.65^{+0.01}_{-0.02}$ & $0.70^{+0.04}_{-0.07}$ & $1.22^{+0.04}_{-0.06}$ & $1.37^{+0.11}_{-0.09}$\\  \smallskip 
 $\Rtwohc$  & $1.00$ & $1.00$ & $1.00$ & $1.17^{+0.07}_{-0.04}$\\  \smallskip 
 $\Rfivehm$  & $1.11^{+0.01}_{-0.01}$ & $1.06^{+0.03}_{-0.02}$ & $0.96^{+0.01}_{-0.01}$ & $1.14^{+0.06}_{-0.04}$\\  \smallskip 
 $\Rvir$ & $1.34^{+0.03}_{-0.02}$ & -- & $0.92^{+0.03}_{-0.03}$ & $1.11^{+0.06}_{-0.04}$\\  \smallskip 
 $\Rtwohm$ & $1.66^{+0.06}_{-0.04}$ & -- & $0.90^{+0.04}_{-0.04}$ & $1.10^{+0.06}_{-0.04}$\\
 \hline
\end{tabular}
\label{apptab:R_err_sample}
\end{table}
\begin{table}
\caption{The two parameter fits [Eq.~\eqref{eqn:fit_format}] within $\Rfivehc$ and a measure of the intrinsic scatter [Eq.~\eqref{eqn:sigma_def}] in the fits at $z=0$ for the median and 16 and 84 percentiles of the cross simulation averaged sample.} \centering
\begin{tabular}{lccc}
 median & $A$ & $B$ & $\slog$ \\ 
 \hline \smallskip
$\Ty$  & $2.048^{+0.003}_{-0.003}$ & $0.575^{+0.001}_{-0.002}$ & $0.1020^{+0.0002}_{-0.0002}$\\ \smallskip
$\Tm$  & $1.763^{+0.002}_{-0.002}$ & $0.576^{+0.001}_{-0.001}$ & $0.0790^{+0.0000}_{-0.0000}$\\ \smallskip
$\Tsl$ & $1.570^{+0.003}_{-0.004}$ & $0.565^{+0.003}_{-0.003}$ & $0.2050^{+0.0007}_{-0.0007}$\\
 \hline
 16\% &&& \\
 \hline \smallskip
 $\Ty$  & $1.770^{+0.003}_{-0.003}$ & $0.584^{+0.002}_{-0.003}$ & --\\ \smallskip
 $\Tm$  & $1.558^{+0.003}_{-0.002}$ & $0.588^{+0.002}_{-0.002}$ & --\\ \smallskip
 $\Tsl$ & $1.255^{+0.006}_{-0.005}$ & $0.616^{+0.012}_{-0.006}$ & --\\
 \hline
 84\% &&& \\
 \hline \smallskip
 $\Ty$  & $2.370^{+0.006}_{-0.006}$ & $0.563^{+0.005}_{-0.004}$ & --\\ \smallskip
 $\Tm$  & $2.005^{+0.007}_{-0.004}$ & $0.570^{+0.010}_{-0.003}$ & --\\ \smallskip
 $\Tsl$ & $1.869^{+0.006}_{-0.008}$ & $0.593^{+0.008}_{-0.011}$ & --\\
 \hline
\end{tabular}
\label{apptab:fits_R5_median_z0_all}
\end{table}
\begin{table}
\caption{The three parameter fits [Eq.~\eqref{eqn:fit_format}] within $\Rfivehc$  and a measure of the intrinsic scatter [Eq.~\eqref{eqn:sigma_def}] in the fits at $z=0$ for the median and 16 and 84 percentiles within the cross simulation averaged sample.} \centering
\begin{tabular}{lcccc}
 median & $A$ & $B$ & $C$ & $\slog$ \\ 
 \hline \smallskip
 $\Ty$  & $1.996^{+0.005}_{-0.005}$ & $0.557^{+0.001}_{-0.001}$ & $0.023^{+0.001}_{-0.002}$ & $0.1007^{+0.0000}_{-0.0000}$\\ \smallskip
 $\Tm$  & $1.772^{+0.004}_{-0.004}$ & $0.579^{+0.001}_{-0.001}$ & $-0.004^{+0.001}_{-0.001}$ & $0.0790^{+0.0000}_{-0.0000}$\\ \smallskip
 $\Tsl$ & $1.736^{+0.006}_{-0.005}$ & $0.633^{+0.002}_{-0.002}$ & $-0.089^{+0.003}_{-0.003}$ & $0.1886^{+0.0001}_{-0.0001}$\\
 \hline
 16\% &&& \\
 \hline \smallskip
 $\Ty$  & $1.706^{+0.008}_{-0.008}$ & $0.564^{+0.001}_{-0.001}$ & $0.034^{+0.004}_{-0.005}$ & --\\ \smallskip
 $\Tm$  & $1.552^{+0.005}_{-0.006}$ & $0.586^{+0.001}_{-0.001}$ & $0.003^{+0.004}_{-0.003}$ & --\\ \smallskip
 $\Tsl$ & $1.414^{+0.014}_{-0.023}$ & $0.681^{+0.002}_{-0.002}$ & $-0.111^{+0.022}_{-0.010}$ & --\\
 \hline
 84\% &&& \\
 \hline \smallskip
 $\Ty$  & $2.292^{+0.014}_{-0.016}$ & $0.545^{+0.001}_{-0.001}$ & $0.031^{+0.008}_{-0.006}$ & --\\ \smallskip
 $\Tm$  & $1.988^{+0.010}_{-0.027}$ & $0.566^{+0.001}_{-0.001}$ & $0.007^{+0.017}_{-0.005}$ & --\\ \smallskip
 $\Tsl$ & $1.911^{+0.030}_{-0.020}$ & $0.604^{+0.001}_{-0.001}$ & $-0.020^{+0.012}_{-0.020}$ & --\\
 \hline 
\end{tabular}
\label{apptab:fits_3_R5_median_z0_all}
\end{table}

\begin{table}
\caption{The two and three parameter $T$-$Y$ fits [Eq.~\eqref{eqn:Yfit_format}] and a measure of the intrinsic scatter [Eq.~\eqref{eqn:sigma_def}] in the fits at $z=0$ for the median and 16 and 84 percentiles of the cross simulation averaged sample within $\Rtwohc$.} \centering
\begin{tabular}{lcccc}
 median & $A$ & $B$ & $C$ & $\slog$ \\ 
 \hline \smallskip
 $\Ty$  & $2.887^{+0.009}_{-0.035}$ & $0.323^{+0.001}_{-0.005}$ & 0 & $0.1154^{+0.0003}_{-0.0013}$\\ \smallskip
 $\Ty$  & $2.525^{+0.011}_{-0.008}$ & $0.335^{+0.002}_{-0.005}$ & $0.025^{+0.001}_{-0.003}$ & $0.1069^{+0.0001}_{-0.0001}$\\ \smallskip
 $\Tm$  & $2.395^{+0.015}_{-0.020}$ & $0.326^{+0.002}_{-0.003}$ & 0 & $0.0913^{+0.0005}_{-0.0007}$\\ \smallskip
 $\Tm$  & $2.146^{+0.008}_{-0.008}$ & $0.336^{+0.002}_{-0.004}$ & $0.021^{+0.001}_{-0.002}$ & $0.0860^{+0.0001}_{-0.0001}$\\
 \hline
 16\% &&& \\
 \hline \smallskip
 $\Ty$  & $2.340^{+0.010}_{-0.009}$ & $0.318^{+0.001}_{-0.001}$ & 0 & --\\ \smallskip
 $\Ty$  & $2.123^{+0.008}_{-0.008}$ & $0.341^{+0.002}_{-0.002}$ & $0.024^{+0.001}_{-0.001}$ & --\\ \smallskip
 $\Tm$  & $1.993^{+0.006}_{-0.006}$ & $0.324^{+0.001}_{-0.001}$ & 0 & --\\ \smallskip
 $\Tm$  & $1.846^{+0.005}_{-0.004}$ & $0.342^{+0.002}_{-0.001}$ & $0.019^{+0.001}_{-0.001}$ & --\\
 \hline
 84\% &&& \\
 \hline \smallskip
 $\Ty$  & $3.321^{+0.011}_{-0.014}$ & $0.305^{+0.001}_{-0.001}$ & 0 & --\\ \smallskip
 $\Ty$  & $3.013^{+0.010}_{-0.009}$ & $0.328^{+0.001}_{-0.002}$ & $0.024^{+0.001}_{-0.001}$ & --\\ \smallskip
 $\Tm$  & $2.729^{+0.012}_{-0.009}$ & $0.312^{+0.001}_{-0.001}$ & 0 & --\\ \smallskip
 $\Tm$  & $2.539^{+0.006}_{-0.006}$ & $0.329^{+0.002}_{-0.001}$ & $0.018^{+0.001}_{-0.001}$ & --\\
 \hline
\end{tabular}
\label{apptab:fits_Y_percentiles_z0_all}
\end{table}
\begin{table}
\caption{The two  parameter $\Ty$-$Y$ fits [Eq.~\eqref{eqn:Yfit_format}] for each sample and a measure of the intrinsic scatter [Eq.~\eqref{eqn:sigma_def}] in the fits at $z=0$ for the median within $\Rtwohc$.} \centering
\begin{tabular}{lcccc}
 median & $A$ & $B$ & $C$ & $\slog$ \\ 
 \hline \smallskip
 $\Bah$+ & $3.053^{+0.013}_{-0.036}$ & $0.313^{+0.001}_{-0.004}$ & 0 & $0.0979^{+0.0005}_{-0.0023}$ \\ \smallskip
 $\Bah$+ & $2.690^{+0.009}_{-0.008}$ & $0.323^{+0.001}_{-0.001}$ & $0.023^{+0.001}_{-0.001}$ & $0.0815^{+0.0001}_{-0.0001}$ \\ 
 \hline \smallskip
 $\The$  & $2.411^{+0.012}_{-0.011}$ & $0.338^{+0.001}_{-0.001}$ & 0 & $0.1452^{+0.0003}_{-0.0003}$ \\ \smallskip
 $\The$  & $2.294^{+0.008}_{-0.009}$ & $0.350^{+0.003}_{-0.002}$ & $0.013^{+0.001}_{-0.001}$ & $0.1450^{+0.0003}_{-0.0003}$ \\
 \hline \smallskip
 $\Mag$  & $2.903^{+0.057}_{-0.052}$ & $0.329^{+0.006}_{-0.005}$ & 0 & $0.0462^{+0.0010}_{-0.0006}$ \\ \smallskip
 $\Mag$  & $2.789^{+0.024}_{-0.027}$ & $0.379^{+0.006}_{-0.008}$ & $0.030^{+0.003}_{-0.003}$ & $0.0461^{+0.0015}_{-0.0009}$ \\
 \hline \smallskip
 $\Tng$  & $2.326^{+0.028}_{-0.040}$ & $0.316^{+0.004}_{-0.005}$ & 0 &  $0.0464^{+0.0011}_{-0.0013}$\\\smallskip
 $\Tng$  & $2.154^{+0.020}_{-0.022}$ & $0.365^{+0.007}_{-0.008}$ & $0.032^{+0.003}_{-0.003}$ & $0.0353^{+0.0003}_{-0.0003}$ \\
 \hline
\end{tabular}
\label{apptab:fits_Y_median_z0_sample}
\end{table}
\begin{table}
\caption{The two parameter $\Ty$-$Y$ fits [Eq.~\eqref{eqn:Yfit_format}] for the `hot' haloes (that is, $Y_{\rm 200c} > 10^{-6}$) and a measure of the intrinsic scatter [Eq.~\eqref{eqn:sigma_def}] in the fits across redshifts for the median and 16 and 84 percentiles of the cross simulation averaged sample within $\Rtwohc$.} \centering
\begin{tabular}{lccc}
 $z=0.00$ & $A$ & $B$ & $\slog$ \\ 
 \hline \smallskip
 50\%  & $2.614^{+0.006}_{-0.006}$ & $0.368^{+0.003}_{-0.010}$ & $0.1873^{+0.0041}_{-0.0152}$\\ \smallskip
 16\%  & $2.216^{+0.008}_{-0.008}$ & $0.357^{+0.003}_{-0.003}$ & -- \\ \smallskip
 84\%  & $3.102^{+0.009}_{-0.009}$ & $0.350^{+0.002}_{-0.003}$ & -- \\ 
 \hline
 $z=0.25$ &&& \\
 \hline \smallskip
 50\%  & $2.593^{+0.007}_{-0.007}$ & $0.372^{+0.002}_{-0.005}$ & $0.1685^{+0.0036}_{-0.0074}$\\ \smallskip
 16\%  & $2.193^{+0.008}_{-0.007}$ & $0.367^{+0.003}_{-0.003}$ & -- \\ \smallskip
 84\%  & $3.093^{+0.010}_{-0.010}$ & $0.355^{+0.002}_{-0.002}$ & -- \\ 
 \hline
 $z=0.50$ &&& \\
 \hline \smallskip
 50\%  & $2.593^{+0.008}_{-0.008}$ & $0.381^{+0.003}_{-0.002}$ & $0.1625^{+0.0034}_{-0.0033}$ \\ \smallskip
 16\%  & $2.177^{+0.009}_{-0.008}$ & $0.380^{+0.003}_{-0.003}$ & -- \\ \smallskip
 84\%  & $3.074^{+0.010}_{-0.010}$ & $0.354^{+0.003}_{-0.003}$ & -- \\ 
 \hline
 $z=1.00$ &&& \\
 \hline \smallskip
 50\%  & $2.585^{+0.013}_{-0.012}$ & $0.382^{+0.004}_{-0.004}$ & $0.1388^{+0.0044}_{-0.0045}$ \\ \smallskip
 16\%  & $2.145^{+0.018}_{-0.020}$ & $0.376^{+0.008}_{-0.014}$ & -- \\ \smallskip
 84\%  & $3.091^{+0.020}_{-0.023}$ & $0.361^{+0.008}_{-0.010}$ & -- \\ 
 \hline
 $z=1.50$ &&& \\
 \hline \smallskip
 50\%  & $2.597^{+0.052}_{-0.031}$ & $0.373^{+0.020}_{-0.011}$ & $0.1263^{+0.0185}_{-0.0091}$ \\ \smallskip
 16\%  & $2.228^{+0.024}_{-0.023}$ & $0.400^{+0.009}_{-0.008}$ & -- \\ \smallskip
 84\%  & $3.115^{+0.032}_{-0.045}$ & $0.368^{+0.008}_{-0.014}$ & -- \\ 
 \hline
\end{tabular}
\label{apptab:fits_Yhot_percentiles_redshift_all}
\end{table}
Here we tabulate a number of the data fits referenced in the text above. In general, the haloes have been processed by taking the data into logarithmically spaced mass (or $Y$) bins, and calculating the median, 16 and 84 percentiles within these bins. The fits are then the fits to these data points, with respect to the median mass (or $Y$) within the same bin.

In particular, the fits we have tabulated here are: Tables~\ref{apptab:fits_median_z0_all}
and \ref{apptab:fits_3_median_z0_all}, the 2- and 3-parameter cross-simulation sample $T-M$ fits at $z=0$, including the 16 and 84 percentiles; Table~\ref{apptab:fits_median_z0_samples}, the 2- and 3-parameter median $T-M$ fits for each sample individually at $z=0$; Table~\ref{apptab:fits_median_Redshift_all}, the 2-parameter median $T-M$ fits for the cross-simulation sample at each redshift; Table~\ref{apptab:R_err_sample}, the shifts and errors in the radius, mass and temperatures over varying radial halo extent definitions for each sample and the cross-simulation sample at $z=0$; Tables~\ref{apptab:fits_R5_median_z0_all} and \ref{apptab:fits_3_R5_median_z0_all}, the 2- and 3-parameter cross-simulation sample, within $\Rfivehc$, $T-M$ fits at $z=0$, including the 16 and 84 percentiles; Tables~\ref{apptab:fits_Y_percentiles_z0_all} and \ref{apptab:fits_Y_median_z0_sample}, the 2- and 3-parameter $T-Y$ fits for the cross-simulation sample and each individual sample respectively; and Table~\ref{apptab:fits_Yhot_percentiles_redshift_all}, the 2-parameter $\Ty-Y$ fits over each redshift for the cross-simulation sample, taking only haloes with $Y_{\rm 200c}>10^{-6}$, including the 16 and 84 percentiles.

\bsp
\label{lastpage}
\end{document}